\documentclass[draft]{agujournal2019}
\usepackage{url} 
\usepackage[inline]{trackchanges} 
\usepackage{soul}
\usepackage{amsmath}
\draftfalse

\journalname{JGR: Machine Learning and Computation}

\begin{document}

%
%


\title{Clustering of Global Magnetospheric Observations}

%
%




\authors{James Edmond\affil{1},
         Joachim Raeder\affil{1},
         Banafsheh Ferdousi\affil{2},
         Matthew Argall\affil{1},
         Maria Elena Innocenti\affil{3}}


\affiliation{1}{University of New Hampshire, Durham, NH, USA}
\affiliation{2}{Air Force Research Laboratory, Albuquerque, NM, USA}
\affiliation{3}{Ruhr-Universität Bochum, Bochum, NRW, Germany}




\correspondingauthor{James Edmond}{james.edmond@unh.edu}




\begin{keypoints}
\item Unsupervised methods are used to cluster MMS and THEMIS measurements into magnetosphere, magnetosheath, and solar wind observations.
\item Combined unsupervised methods can broaden clustering algorithms, with hierarchical clustering capturing magnetospheric population diversity.
\item Using the predictions of our model, we created a dataset of 6131 magnetopause and 3238 bow shock crossings.
\end{keypoints}

%
%

%
%


\begin{abstract}
The use of supervised methods in space science have demonstrated powerful capability in classification tasks, but unsupervised methods have been less utilized for the clustering of spacecraft observations. We use a combination of unsupervised methods, being principal component analysis, self-organizing maps, and hierarchical agglomerative clustering, to make predictions on if THEMIS and MMS observations occurred in the magnetosphere, magnetosheath, or the solar wind. The resulting predictions are validated visually by analyzing the distribution of predictions and studying individual time series. Particular nodes in the self organizing map are studied to see what data they represent. The capability of deeper hierarchical analysis using this model is briefly explored. Finally, the changes in region prediction can be used to infer magnetopause and bow shock crossings, which can act as an additional method of validation, and are saved for their utility in solar wind validation, understanding magnetopause processes, and the potential to develop a bow shock model.

\end{abstract}

\section*{Plain Language Summary}
Machine learning in space science often uses supervised methods for classification, but we explore using unsupervised methods for clustering spacecraft observations. We combine principal component analysis, self-organizing maps, and hierarchical clustering to predict whether observations occurred in the magnetosphere, magnetosheath, or solar wind for THEMIS and MMS. We visually validate predictions, study specific map nodes for data representation, and briefly discuss deeper hierarchical analysis. Additionally, we use region prediction changes to identify magnetopause and bow shock crossings.

\section{Introduction}

The region of space where Earth is directly affected by solar activity can be divided into various regions, such as the solar wind, the magnetosheath, and the magnetosphere itself. Since the first measurements of one such region, the solar wind, were made incidentally by \citeA{first_SW_measurement_Soviet} and intentionally by \citeA{first_SW_measurement_American}, more than dozens of missions have recorded measurements in these different regions to investigate various space plasma processes. The solar wind is a continual stream of plasma ejecta originating from the Sun that is accelerated in the solar corona, although the exact mechanisms through which it does this have not yet been confirmed \cite{solar_corona}. This incident plasma is slowed to subsonic speeds by Earth's magnetosphere, resulting in a piled-up region of heated and slowed plasma called the magnetosheath \cite{magsheath}. The boundary layer that defines the separation of these two regions is called the bow shock, which has a stand-off distance of approximately 14 Earth radii along the x axis and is generally modeled as a hyperbola in the x-y plane in GSE coordinates, although its shape and position can vary in response to solar wind parameters \cite{Fairfield_LowMachNumberShocks}. The magnetosphere itself can be further compartmentalized into a number of sub-regions containing plasma with different properties, including the ring current, the radiation belts, the ionosphere, the tail, etc. The boundary layer separating this region from the magnetosheath is the magnetopause, which acts as the pressure equilibrium between the magnetic pressure of the earth and the dynamic pressure of the shocked solar wind \cite{magnetopause}.

In analyzing the measurements of spacecraft that frequent these regions, it is generally not difficult to identify which region a spacecraft is in given a brief time history. The same can often be said for doing so with a joint set of measurements at a single moment in time. Drafting general mathematical relationships that can always predict the regions is more challenging. Classification in the context of machine learning is ideal for this task as it involves predicting what class a data point belongs to. This requires that one curates a dataset and provides the class label for each point. There have already been many successful efforts in using this approach for different missions, such as \citeA{mms_region_classification}, \citeA{argall_mms_sitl_ml}, and \citeA{mms_cnn_olshevsky} using deep learning or \citeA{nguyen_ml_method}, \citeA{ssc_ml}, and \citeA{sw_categories_supervised_ml} for more traditional nonlinear approaches. Clustering is an unsupervised task in which data are amalgamated into homogeneous groups and more recently, some have used related methods to predict classes of data, like \citeA{SW_AEandGSOM} with predicting solar wind classifications from ACE data, \citeA{som_openggcm} with predicting different regions in magnetospheric simulation results, and \citeA{fully_kinetic_sim_SOM} for classifying PIC simulations involving the tearing instability.

Our methodology is to use an unsupervised approach to separate the solar wind, magnetosheath, and magnetosphere measurements from spacecraft data. Our data are recorded at different time resolutions, so methods reliant upon a consistent time step cannot be utilized and must focus on the joint set of measurements alone. We use principal component analysis to reduce the dimensionality and correlations in our dataset, along with a visualization technique to add greater interpretability to the dimensionality reduction. Self-Organizing Maps (SOMs) \cite{Kohonen_original_som} are then used to effectively reduce the size of the training set so that a larger number of clustering algorithms can be considered. We finally use hierarchical agglomerative clustering to cluster the individual nodes of the SOM and propagate the cluster assignments of the nodes to the data they represent (an overview of various hierarchical clustering methods is covered in \citeA{hierarchical_clustering_overview} and performance metrics for specifically hierarchical agglomerative clustering methods in \citeA{hierarch_agg_clustering_methods}). The use of hierarchical clustering coupled with a self-organizing map provides a unique advantage in that in addition to being able to separate the data into clusters, these clusters are composed of subclusters which can be further investigated. This combination distinguishes it from other more common clustering methods. The paper is outlined as follows: data sources, data preprocessing, dimensionality reduction, self-organizing maps, clustering of SOM nodes, results, boundary crossings, and conclusion.




\section{Data Sources}

We use data from two missions, Time History of Events and Macroscale Interactions during Substorms, \cite{Angelopoulos_THEMIS}, or THEMIS, and the Magnetospheric Multiscale Mission \cite{MMS}, or MMS. This data includes measurements of magnetic field $\mathbf{B}$, the ion velocity $\mathbf{V}$, the ion temperature $T$, and the ion density $n$, a cumulative eight features. Below, we describe for each mission how the data is prepared.



\subsection{THEMIS}

THEMIS is a collection of five spacecraft (THEMIS-A, B, C, D, and E) that orbit in near-Earth space. We used data from March 2007 to the end of December 2021. THEMIS-B and C were moved to lunar orbit in 2009 to become the Acceleration, Reconnection, Turbulence, and Electrodynamics of the Moon's Interaction with the Sun \cite{Angelopoulos_ARTEMIS} (ARTEMIS) mission where they would make measurements departing from what would normally be seen by THEMIS-A D, and E. We only use THEMIS-B and C data up until end of year 2009.

The ion velocity, temperature, and density measurements of THEMIS are from the Electrostatic Analyzer instrument \cite{McFadden} and are available at multiple time resolutions, such as ``reduced" (ESAR) and ``full" (ESAF) data packets. The ESAR offers higher time resolution at once per spin ($\sim$3 secs), but the cold temperatures of typical solar wind mean that their distributions are narrow and require sufficiently high angular resolution to resolve. The ESAF packets sacrifice time resolution for higher angular resolution and are available in two formats, 32-spin (96 sec) in fast survey mode and 128-spin ($\sim$6.5 minutes) in slow survey mode. Figure 5 of \citeA{McFadden} illustrates the difference in angular resolution. The data are flagged for quality and we use quality zero data, indicating no issues. The magnetic field measurements are from the Flux Gate Magnetometer (FGM) \cite{THEMIS_FGM} and are collected at spin resolution. This data is then averaged down to the resolution of the ESAF measurements (e.g. 6.5 mins for fast survey and 96 secs for slow) to synchronize them. 




\subsection{MMS}

MMS, the Magnetospheric Multiscale Mission \cite{MMS}, is a constellation of four spacecraft (MMS-1, 2, 3, and 4) flying in equatorial orbits in mutual close proximity to make electron-scale measurements. The ion measurements are taken from the Dual Ion Spectrometer as part of the Fast Plasma Investigation (FPI) \cite{MMS_FPI}. Multiple ion spectrometers per spacecraft makes it possible to make measurements below spin resolution. The magnetic field measurements are taken from the Flux Gate Magnetometer (FGM) \cite{MMS_FGM} and are available at 10 ms. These magnetic field measurements and ion measurements are averaged down together to 1 minute resolution. Data from MMS 1, 2, and 3 span September 2015 to December 2021. Due to damage to the spectrometers of MMS 4, we only use data from September 2015 to 7 June 2018.







\subsection{Data Cleaning}









The THEMIS and MMS datasets possess 8.13 and 4.09 milion points, respectively. The methods we apply to these data can be very sensitive to outliers and the size of magnetic field measurements closer to Earth could impact our ability to separate them in an unsupervised manner. We constrain our data to be between 7 and 35 Earth radii. This final filtering leaves us with 9.64 million points. We separate our data with a test-train split of 95\% to 5\%, giving us a training size of $\sim$480k points.


\section{Data Preprocessing}



\begin{figure}
\noindent\includegraphics[width=\textwidth]{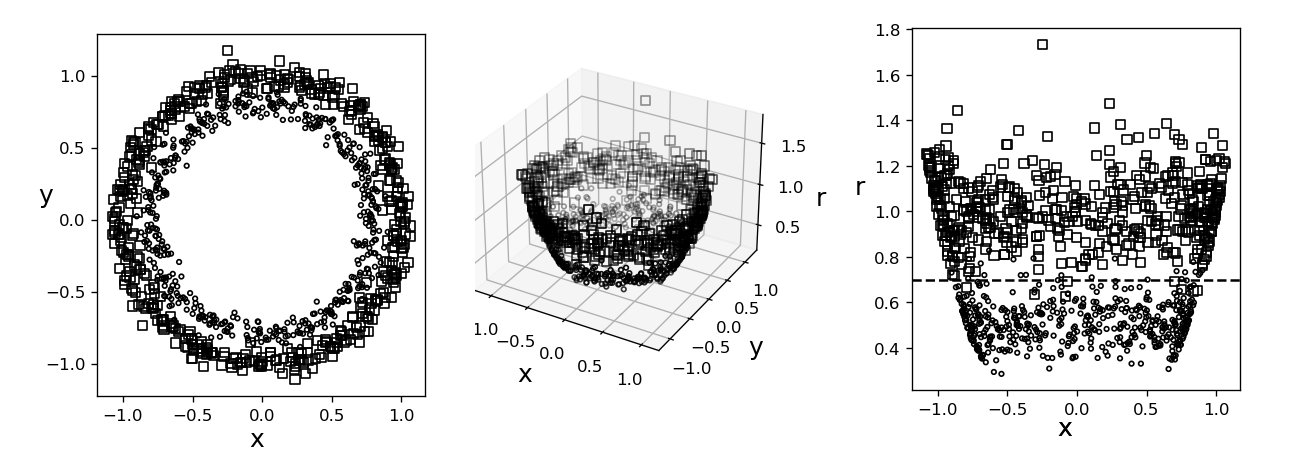}
\caption{A basic example demonstrating the importance of feature engineering. Two noisy concentric circles (plotted in the left figure with different markers) cannot be separated from each other linearly according to their x and y positions alone. Creating the feature $r = \sqrt{x^2 + y^2}$ and plotting (x,y,r) shows a correlation of one circle with higher values of r. This is seen more clearly in a plot of (x,r) in the right figure where a line can separate the circles to high accuracy.}
\label{nonlinear_feature_example}
\end{figure}

The eight features, \textbf{V}, \textbf{B}, $n$, and $T$, of our dataset do not possess enough variance for many unsupervised methods to sufficiently separate the regions. It is very common within machine learning to engineer derived features from the original in hopes of capturing non-linear relationships \cite{autofeat}. This is because what is non-linearly separable in some space might become linearly separable in a higher dimensional space (see the example in Figure \ref{nonlinear_feature_example}). In addition to the vector components of \textbf{V} and \textbf{B}, we include the ion speed $V$ and the magnetic field magnitude $B$ as two extra features. We also create the ion momentum density $\mathbf{mom} = n \mathbf{V}$ with ion mass set to 1 as three additional features, giving us a total of 13 features.


\begin{figure}
\noindent\includegraphics[width=\textwidth]{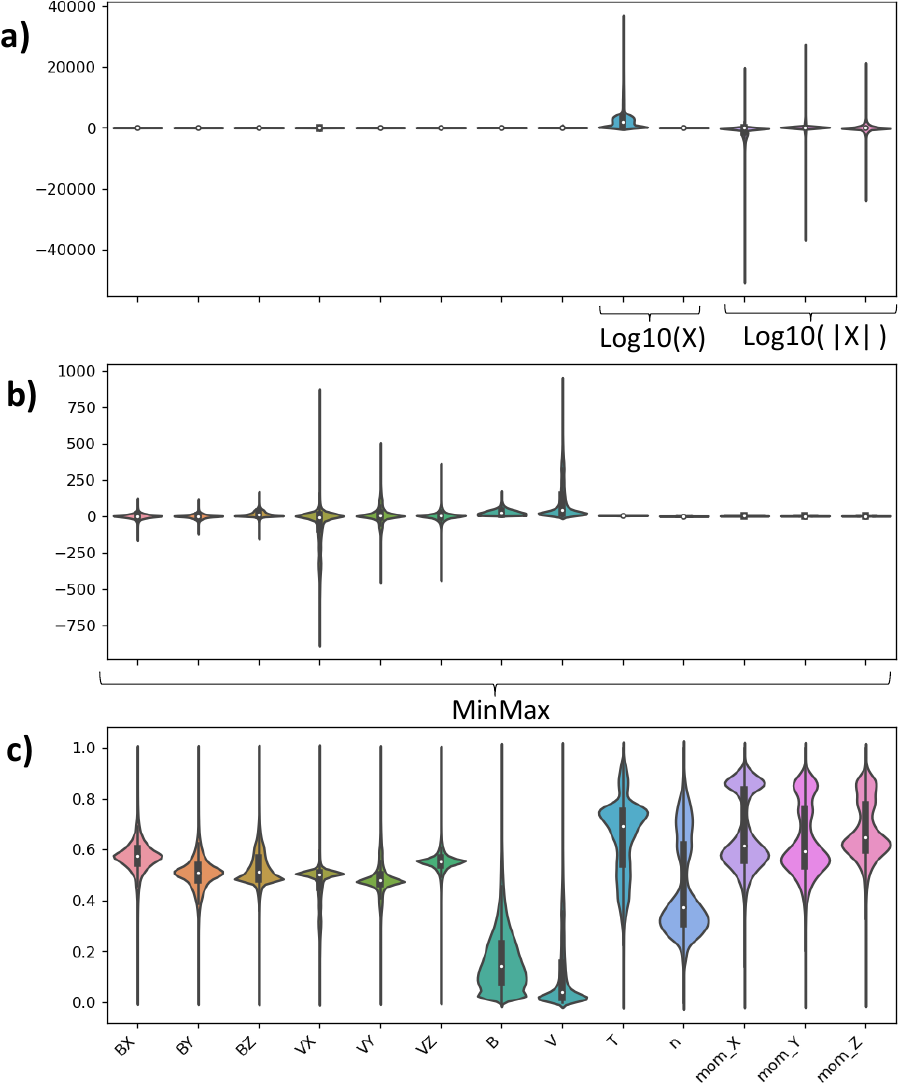}
\caption{A series of violinplots representing the distributions of features as we apply different transformations. A violinplot shows the kernel density estimate as the width, the range of the estimate as a thin vertical grey bar, the interquartile range as a thick vertical black bar, and the median as a white dot. Plot a shows the original range per feature, b shows the log10-transformed values (log10 of $n$ and $T$, log10-absolute of momentum density), and c shows the min-max scaling of the results of b.}
\label{violinplots}
\end{figure}

Most of the features have ranges over a few hundreds whereas the density, temperature, and momentum density components cover multiple orders of magnitude. We convert the density and temperature to log10 scale, but the negative values of the momentum density will not allow for this transform. We will transform the momentum density using the log10 of the absolute values of their components instead. After, these data still possess uneven ranges. For the dimensionality reduction and later clustering methods we use, large magnitudes can have a significant impact on the results. Common scalings that are used can include standardization to modify the feature of each variable to mean 0 and unit variance, min-max normalization to change the minimum and maximum of each feature to 0 and 1, respectively, or certain power transforms (e.g. to make data more gaussian-like using the Box-Cox transform \cite{box_cox}). We use a min-max rescaling that is fitted on the training data to bring all features into a similar range. The pipeline of data transformations performed is shown in Figure \ref{violinplots}.




\begin{figure}
\noindent\includegraphics[width=\textwidth]{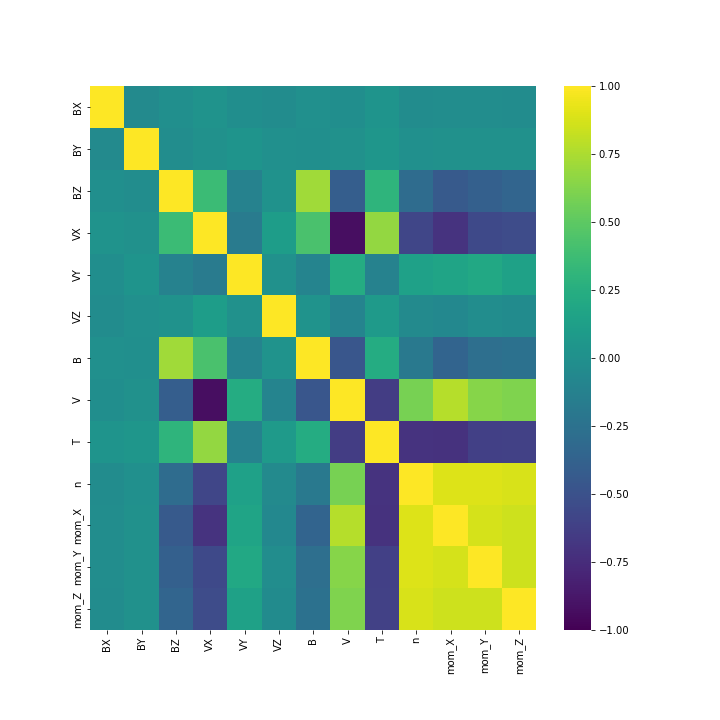}
\caption{A heatmap of the correlations between variables in the MinMax-rescaled training dataset. The plot is symmetric across the diagonal. It is to be interpreted as showing the correlation of each feature with every other feature in the training set, e.g. correlation( log10(n), log10(T) ) $\sim$ -0.6, or the log10 of the density is moderately negatively correlated with log10 of the temperature. There is a visible number of variable pairs with large magnitude in correlation (the bright or dark colored boxes in the off-diagonal). Also apparent is the absence of correlation of $BX$ and $BY$ with all other variables - even with B. The lack of correlation with the B is due to $BX$ and $BY$ being symmetric distributions around 0. The $VX$ and $VY$ components of $\mathbf{V}$ have correlation with V because large speeds ($>$400 km/s) are often going to be associated with solar wind (generally possessing large negative magnitudes in VX) and positive VY is also going to be more associated with solar wind due to the angle it arrives at the Earth (with an average VY being around 30 km/s). }
\label{featcorr}
\end{figure}

We show in Figure \ref{featcorr} that after these rescalings, the features in our training set are generally correlated. The high number of correlated features means that direct clustering methods would be biased in the favour of these correlated components. Further still, the dimensionality can make some methods expensive to compute or to return poor solutions due to the curse of dimensionality. The implication of the latter here is that distances between points will become smaller as the dimensionality increases, reducing the quality of clustering solutions. For data that does not possess significant outliers or that has been meticulously cleaned, the loss in quality of these solutions may be small, but it can become an issue for noisy data, especially data that are observations. We address both the correlation of features and dimensionality in the dimensionality reduction method to follow.

\section{Dimensionality Reduction}

\begin{figure}
\noindent\includegraphics[width=\textwidth]{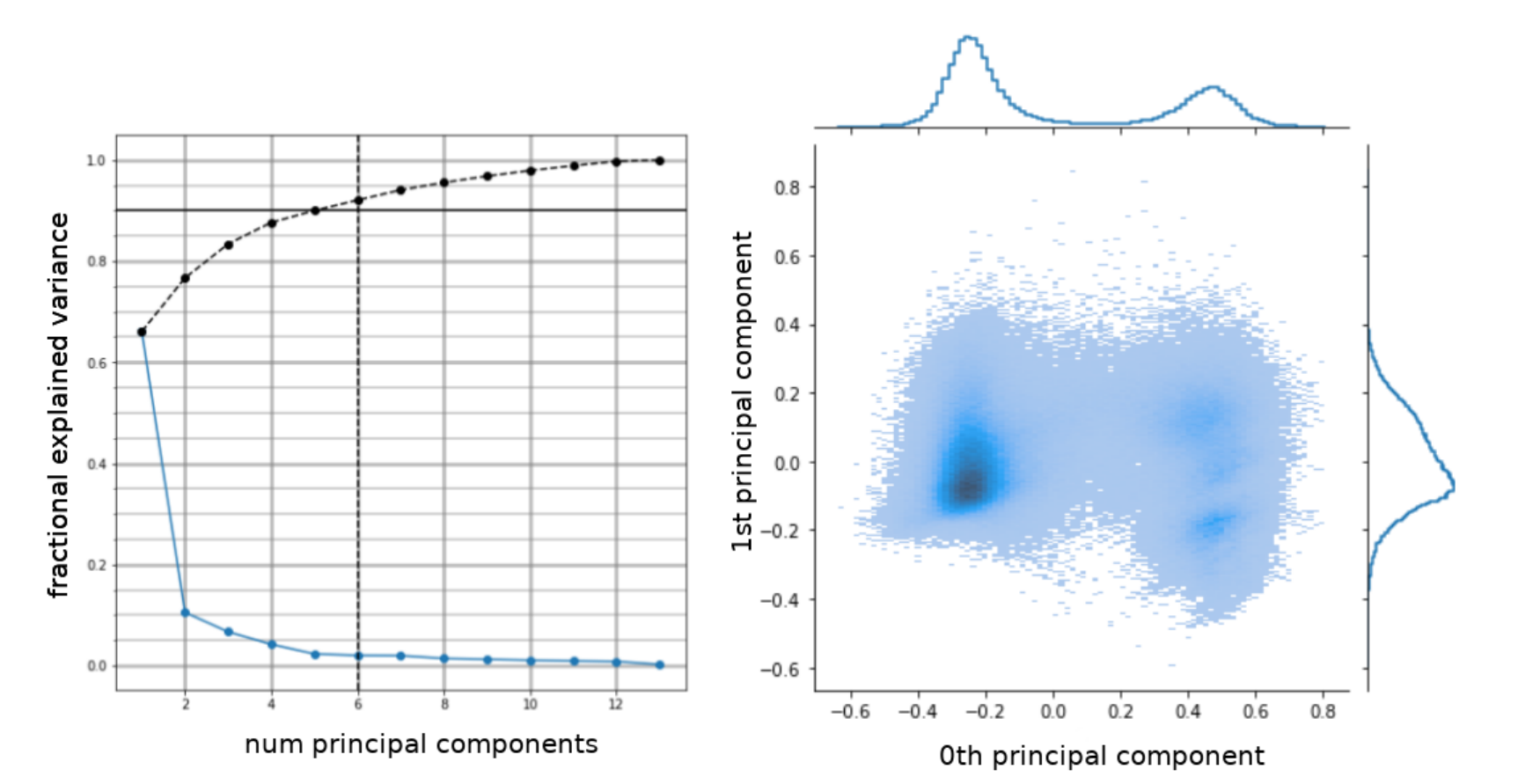}
\caption{ \textbf{Left}: The eigenvalues (normalized relative to their sum) from the PCA decomposition are plotted in descending order as the solid blue line. The cumulative sum of these normalized eigenvalues is plotted as the upward-trending dashed black line. We choose to select a number of components representing at least 90\% of the variance (indicated by the horizontal black line), so 6 components are chosen that represent 93\% (the vertical black line). \textbf{Right}: A bivariate histogram of the training data projected onto the first two principal components, representing 76\% variance. The margins show the histograms for the $0^{th}$ and $1^{st}$ components on the top and right, respectively. It is evident from the first two components that several clusters are present in the data. }
\label{pca_comps}
\end{figure}

Our training set prohibits using many clustering methods due to a combination of the training size, the dimensionality, and the presence of correlated features. We can simultaneously reduce the number of dimensions and the number of correlated features using one of the most prominent dimensionality reduction techniques, principal component analysis \cite{pca}, or PCA. The goal of PCA is to find a new set of uncorrelated variables, called the principal components, that capture the maximum variance in the data. These principal components are ordered by the amount of variance they explain, with the first component explaining the most variance and subsequent components explaining less. A common way this is done is by computing the eigenvalues and eigenvectors of the covariance matrix of the data. The eigenvalues quantify the proportions of variance captured by the eigenvectors and these eigenvectors are the principal components. We plot in descending order the normalized eigenvalues in Figure \ref{pca_comps}. A bivariate histogram of the training data projected onto the first two principal components is also seen in the same figure. PCA has limitations in that it is a linear method of dimensionality reduction. That is, it considers the relationships between features to be linear, or at least approximately so. When data are characterized by non-linear correlations, this complicated structure can be destroyed in the transformation and cause misinterpretations of the resulting components.

\begin{figure}
\noindent\includegraphics[width=\textwidth]{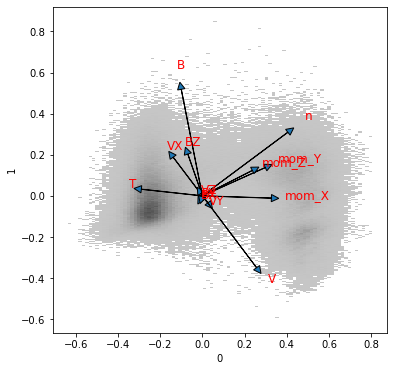}
\caption{ The two-component projected training data is depicted as a bivariate histogram in grayscale. The arrows plotted here are the loadings for our features across the first two principal components. The length of an arrow represents the influence that feature had for the PCA projection along that direction. All arrow lengths are normalized to the longest arrow, that of the B feature. From the plot, the temperature significantly influenced the $0^{th}$ component but barely for the $1^{st}$ and points to the cluster on the left. This means that that cluster is likely to correspond to higher temperatures than the data on the right. The density roughly equally contributed to both components and indicates that the top right region is related to higher densities and by its antiparallel direction, the cluster on the left is largely associated with lower densities. Since VX points to the top left and V to the bottom right, the bottom right region is related to data with high speeds and large negative values of VX. The BX, BY, VY, and VZ features are clustered at the origin, indicating that they did not influence the first two components (although they may have impacted the higher order components). Overall, we can surmise from this plot alone that the left, top right, and bottom right areas are associated with higher temperature, higher density, and higher speeds, respectively. Thus, it is likely that these clusters are the magnetosphere, magnetosheath, and solar wind populations. }
\label{pca_biplots}
\end{figure}

PCA uses linear combinations of features to ascertain directions of maximal variance with projections of the form $PCA_i = \sum_a^D z_{ia} F_a$ where $PCA_i$ indicates the $i^{th}$ principal component, $\{F_a\}$ is the set of D features, and $\{z_{ia}\}$ are the weights in the linear combination. These weights are also referred to as loadings. The loadings can be inspected to determine how much a particular feature contributes to a principal component. Using just the first two principal components, we can visualize these loadings as vectors that can visually communicate the importance of each feature in the projection. Plotting these vectors on top of the data projected onto the first two components is called a biplot and is seen in Figure \ref{pca_biplots}. Using biplots to infer information from PCA results has a rich history and an introduction to the concept is covered in \citeA{pca_biplot}.


Issues of feature correlation and moderate dimensionality are resolved using PCA projections, but there is still the matter of a large training size after the PCA transform. The size can be reduced by simply randomly selecting fewer points, but this will only trade variance for sample size. Choosing enough points to represent a similar amount of variance will still require a large population size. We will use a method in which distinct points act as ``representative" of their local distribution such that their amalgamation reflects the distribution of the training set. 


\section{Self Organizing Maps}


\subsection{Theory}

In a higher-dimensional space of N points, finding m ``prototype" points where m $<<$ N while also minimizing some predefined distortion criteria is the main goal of vector quantization \cite{som_and_vq,vector_quantization}. The distortion criteria changes for different methods, but the most common one involves computing the inertia, or the sum of square distances of each point from their closest representative. One of the most popular clustering methods, KMeans \cite{kmeans}, uses this as a convergence criteria.

Succinctly put, KMeans will partition the data into $k$ voronoi-separated clusters where $k$ is pre-specified. To accomplish this, $k$ random points from the data are selected to act as initial cluster centroids. The distances between points in the data and the centroids are computed and points are assigned to the cluster whose centroid they are closest do. The centroid positions are re-computed as the average of all the points in their respective clusters. This procedure continues until the changes in inertia are small or until a max iteration number is reached. These centroids act as the prototype vectors for the clusters. This type of learning is purely competitive in that centroid updates are only affected by the data in their own clusters. It will have limited success with data that do not contain spherically separable clusters, particularly those non-convex in shape. A work-around to the non-convexity difficulty has been to use KMeans on data to resolve many clusters (often several hundred or more) and then apply a more resilient clustering method to the cluster centroids and propagate the predictions of this second stage clustering to the data represented by the centroids. However, this method will still be subject to the competitive learning biases inherent in KMeans solutions.

\begin{figure}
\centering
\noindent\includegraphics[width=0.6\textwidth]{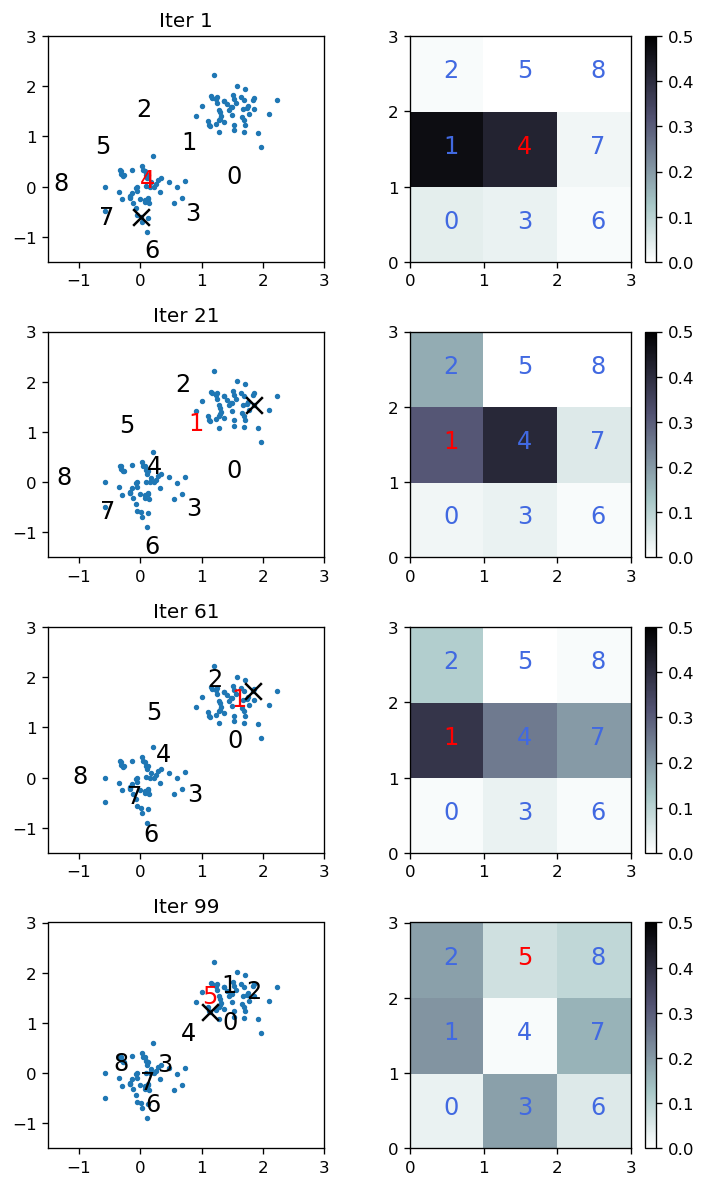}
\caption{ These plots show the convergence of a SOM over iterations on a simple 2d dataset consisting of two isotropic gaussians. \textbf{Left}: The node positions per iteration in the data are shown. The positions are initialized over the first two principal components of the data. For clarity, the positions are plotted as their number in the SOM grid (node 3 is depicted as a `3'). At each iteration, one data point is selected to train the network against (plotted as X) and the BMU for that point is identified and shown in red. \textbf{Right}: The 2d SOM grid is shown per iteration. Heatmaps show the fraction of data points that any node is closest to. At iteration 1, nodes 1 and 4 are the closest nodes to (or, ``represent'') about 90\% of the data. By the final iteration, the data representation is more equidistributed across the network. The distribution could be further improved with a better choice of hyperparameters for this SOM. }
\label{som_example}
\end{figure}

A much more robust method is the Self-Organizing Map, or SOM, which uses a combination of competitive and cooperative updates. A SOM is a method of clustering that is meant to resemble the structure of a neural network. The neurons are referred to as nodes and they are usually arranged in a square 2d grid (i.e. the ``node-space"). Each node has a weight vector $\mathbf{w}$ which is the position of the node relative to the data (i.e. in ``feature-space"). The relationship between higher-dimensional data in the feature-space and the 2d grid of the node-space allows for 2d visualizations of higher dimensional data. To train a SOM, a data point $\mathbf{q}$ is presented to the network and the closest node in feature-space, called the best-matching-unit or BMU, is identified. The BMU will then be moved closer to $\mathbf{q}$. If we interpret similarity between two points as being related to their proximity, then moving the BMU closer to $\mathbf{q}$ can be described as making the BMU more similar to $\mathbf{q}$, or more representative of it. An example of the convergence of a simple SOM on 2d data is shown in Figure \ref{som_example}.

To incorporate a form of Hebbian learning, or ``neurons that fire together wire together," nodes near the BMU are moved closer to $\mathbf{q}$ as well. The amount they are moved is proportional both to their feature-space distance to $\mathbf{q}$ (the distance from the node to $\mathbf{q}$ as seen in the left plots of Figure \ref{som_example}) and their node-space distance to the BMU (i.e. the distance from the node to the BMU as seen in the right plots of Figure \ref{som_example}). This node-space distance is supplied to the neighborhood function and usually involves exponentially diminishing distances. Common neighborhood functions are the gaussian and mexican hat functions, both of which are reliant upon the neighborhood distance hyperparameter $\sigma$ to determine the sharpness of the distribution. The update to the weights of node $a$ per iteration can be expressed with

\begin{linenomath*}
\begin{equation}
\mathbf{w}_{a}(i+1) = \mathbf{w}_{a}(i) + \alpha(i) h(i,a,BMU) (\mathbf{q} - \mathbf{w}_{a}(i))
\end{equation}
\end{linenomath*}

\noindent
where $i$ is the iteration number, $\mathbf{w}_{a}(i)$ is the weight vector for node $a$ at iteration $i$, $\alpha(i)$ is the learning rate at iteration $i$, and $h(i,a,BMU)$ is the neighborhood value between nodes $a$ and the BMU at iteration $i$. The $\mathbf{q} - \mathbf{w}_{a}(i)$ term represents the feature-space distance of node $a$ from $\mathbf{q}$ and the $h(i,a,BMU)$ term is the neighborhood distance of node $a$ from the BMU. Convergence of the SOM is guaranteed by specifying a finite number of iterations. Defining an initial and final $\alpha$ and $\sigma$, a decay function determines the learning rate and neighborhood distance at each iteration $i$. Commonly used decay functions include a linear or exponential decay.

Another aspect of SOMs is the way the nodes are structured relative to each other. In the example SOM used in Figure \ref{som_example}, a node's immediate neighbors are those vertically or horizontally adjacent to it such that node 4 has nodes 1, 5, 3, and 7 as immediate neighbors, node 5 has nodes 2, 4, and 8 as immediate neighbors, and node 0 only has nodes 1 and 3 as immediate neighbors. This structured relationship is referred to as the topology of the network and the type used here is a square topology. Topologies involving any convex shape are technically possible, but only a handful are really utilized, such as the hexagonal topology. If a map with a square topology represents a grid-like structure, then one with a hexagonal topology resembles a style of honeycomb structure. Maps with square topologies are simpler and easier to visualize whereas hexagonal ones can be more difficult to visualize. However, this more complex topology can often more compactly represent data than a square topology and can have less edge effects at the borders of the map. We only consider a square topology for our model.

The quantification of similarity between a SOM and the data is the quantization error Q, given as

\begin{linenomath*}
\begin{equation}
Q = \frac{1}{N} \sum_{i=1}^{N} |x_i - BMU(x_i)|
\end{equation}
\end{linenomath*}

where $x_i$ is the $i^{th}$ point of a dataset of size $N$ and $BMU(x_i)$ is the BMU for point $x_i$. This is simply the concept of inertia described previously, but normalized to the size of the data. A survey of SOM applications and metrics used to verify their accuracy can be found in \citeA{Kohonen_matlab_som_intro}.


\subsection{Implementation}

There are several open-source python packages implementing SOMs available. The most common is minisom \cite{minisom}, which uses a vectorized design to speed up computations. For large datasets or network sizes, the time to completion may still quite long. Traditionally, training a SOM has been a computationally expensive process for two reasons: The network adapts to one point at a time, and it is fairly common that multiple trainings are done. The latter occurs because SOM initialization and training are done stochastically and there is a large number of hyperparameter choices available (the number of iterations, the network size, the decay function, the neighborhood function, the initial and final $\alpha$ and $\sigma$, etc). Since the network with the lowest quantization error is usually selected as the best fitting, this significantly increases the total amount of time needed to get a complete and robust model.

The one-at-a-time training constraint is resolved using  SOMs that train over batch-updates. These usually involve computing weighted averages of the neighborhood values across a batch of samples. This approach is taken by two popular python packages Somoclu \cite{somoclu} and XPySom \cite{xpysom}. Both packages also support CUDA acceleration. It is worth explicitly stating that the speed-up on CPU resources alone is close to a factor of 100, sometimes greater and using GPU resources can push this yet further. We have used the XPySom package for our results.


\subsection{Hyperparameter Optimization and Training}


To expedite the process of finding the best fitting SOM with the most appropriate set of hyperparameters, we create a micro training set. First, we min-max normalize the PCA-projected training data in order to avoid bias to any particular feature. Next, we run KMeans to resolve 10,000 clusters with a kmeans++ initialization method for 100 runs and select the optimal run based on minimal inertia. This initialization method makes better choices for cluster centroids by weighting data in proportion to their square distance from the previously created centroid. Then for each centroid, the closest point in the training data is extracted, and the resulting 10,000 points form the micro training set. The remaining points in the training dataset are referred to as the macro training set with a size of 470k.

We consider a number of different SOM hyperparameters and each SOM will be trained on the micro training set and validated on the macro training set. The maps are validated in this way because the macro set will contain a larger number of outliers and given the noise evident in Figure \ref{pca_biplots}, resolving these outliers correctly will be critical. The hyperparameters of the map with the lowest value for our loss function will be retained and a final SOM will be trained using these hyperparameters on the macro training set. We define our loss function to be

\begin{linenomath*}
\begin{equation}
L = Q \hspace{2pt} * \hspace{2pt} ( \hspace{5pt} \frac{n_x n_y} { (n_x)_{max} (n_y)_{max} } + \frac{max\{n_x,n_y\} } { min\{n_x,n_y\} } \hspace{5pt} ).
\end{equation}
\end{linenomath*}

\noindent
where Q is the quantization error of the SOM, $n_x$ and $n_y$ are the dimensions of the 2d node grid, and $(n_x)_{max}$ and $(n_y)_{max}$ are the maximum values permitted for the x and y dimensions. The $max\{n_x,n_y\} / min\{n_x,n_y\}$ term penalizes non-square networks and will only allow for non-square maps should they provide a sizably lower quantization error.

It should be noted that the use of a custom loss function for SOM validation is critical for our purposes. With the number of training iterations and training data set held constant, increasing the map size will generally reduce the quantization error for many choices of hyperparameters. A larger map size may better represent the training data, and in many cases even the test data, than a smaller map, but a larger number of nodes and their distributions may be suboptimal for clustering methods that will fit to these nodes. This can be loosely seen as a form of overfitting, but not in the sense of a model not generalizing well to unseen data. Rather, as the map size increases and more nodes are pushed to the fringes to represent outliers, the distributions of these nodes become more complex and require more robust clustering methods to resolve correctly. This is why despite a large map producing a better and more detailed 2d visualization of the data, a smaller map size and simpler clustering method (e.g. an 8x8 SOM + KMeans) can counterintuitively cluster complicated data better than a more advanced combination (e.g. a 30x30 SOM + Gaussian Mixture Model).

The python-based optimization library Optuna \cite{optuna} is used to determine which hyperparameter choices are made. The training of each SOM on the micro training set is referred to as a trial. Optuna offers a variety of samplers to generate hyperparameters choices, and we use the Tree-structured Parzen Estimator (TPE) with independent sampling as the sampler. It generates hyperparameter choices by fitting two sets of Gaussian Mixture Models (GMM) per trial, one set for the better performing trials, $l(x)$, and another for the remaining, $g(x)$. Each set involves fitting a GMM for each hyperparameter $x$ and the hyperparameter value selected is that which maximizes the ratio of density estimates $l(x) / g(x)$. Maximizing this ratio is consistent with choosing a hyperparameter that is simultaneously most likely to be generated by $l(x)$ (the ``good" models) and least so by $g(x)$ (the ``poor" models).

For our optimization, we will consider the following hyperparameters. The number of nodes for the SOM grid $n_x$ and $n_y$, the initial learning rate $\alpha$, the initial neighborhood size $\sigma$, the neighborhood function $H$, and the decay function $D$. We have fixed the number of training epochs to be 50, the final learning rate and neighborhood size to be 0.01, and the maximum $n_x$ and $n_y$ dimensions to be 30. The values the hyperparameters are permitted to take are in the following list:

\begin{enumerate}
    \item $ 5 \leq n_x, n_y \leq 30 $
    \item $ 1 \leq \sigma \leq \sqrt{n_x n_y} $
    \item $ 0.1 \leq \alpha \leq 1 $
    \item D: \{linear, exponential\}
    \item H: \{Gaussian, Mexican-Hat\}
\end{enumerate}


\subsection{SOM Results}

After 500 trials, the best hyperparameter options are $(n_x,n_y)$ = (14, 14), $\sigma$ = 5.518, $\alpha$ = 0.843, $D$ = exponential, and $H$ = mexican hat. We train a SOM with these hyperparameters on the macro training set which completes in 7 minutes. The resulting SOM has a quantization error of 0.0702 and 0.0703 on the macro training and test sets. The cost function value for both rounds to 0.0856. With a Intel\textsuperscript{\textregistered{}} Xeon\textsuperscript{\textregistered{}} 2.90GHz E5-2690 (32 cores, 64 threads) CPU and 64 GB of RAM available, the entire process of hyperparameter optimization and final model training takes approximately 3 hours.

\begin{figure}
\noindent\includegraphics[width=\textwidth]{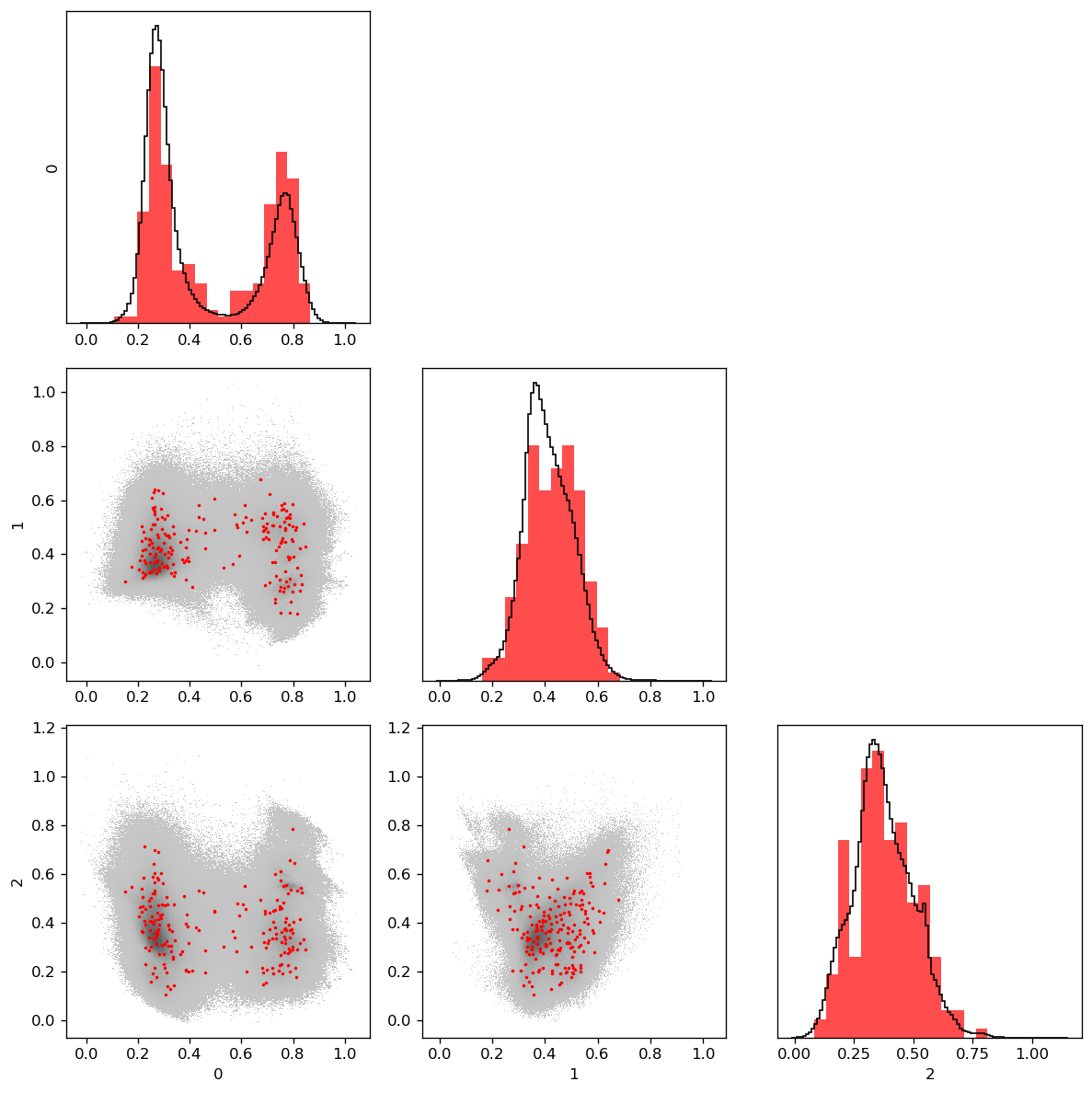}
\caption{ Pairplots over the first 3 min-max normalized principal components (83\% variance) of the test set. The off diagonal plots are bivariate histograms for the test data in greyscale. Scatter plots of the SOM node position are plotted in red on top of the bivariate histograms. The diagonal plots are the marginal distributions where the black line is the test data distributed over 100 bins. The SOM node positions are simultaneously binned but at a smaller resolution of 25 bins. The nodes generally match the histograms of the $0^{th}$ and $2^{nd}$ components with a dip noticeable in the nodes histogram of the $1^{st}$ component. }
\label{som_and_data_pairplot}
\end{figure}

While the SOM we have trained has a good quantization error, there are visualization techniques we can use to give further credence to how well it represents the data. Since the goal of a SOM is to give a vector-quantized representation of the data, one simple approach is to create plots of the data itself with the SOM node positions overlaid. If it is an effective representation, it should roughly map to positions of high data density, both in scatter plots and histogram marginals. We look at pairplots over the first 3 MinMax-normalized principal components of the test set in Figure \ref{som_and_data_pairplot}.

\begin{figure}
\noindent\includegraphics[width=\textwidth]{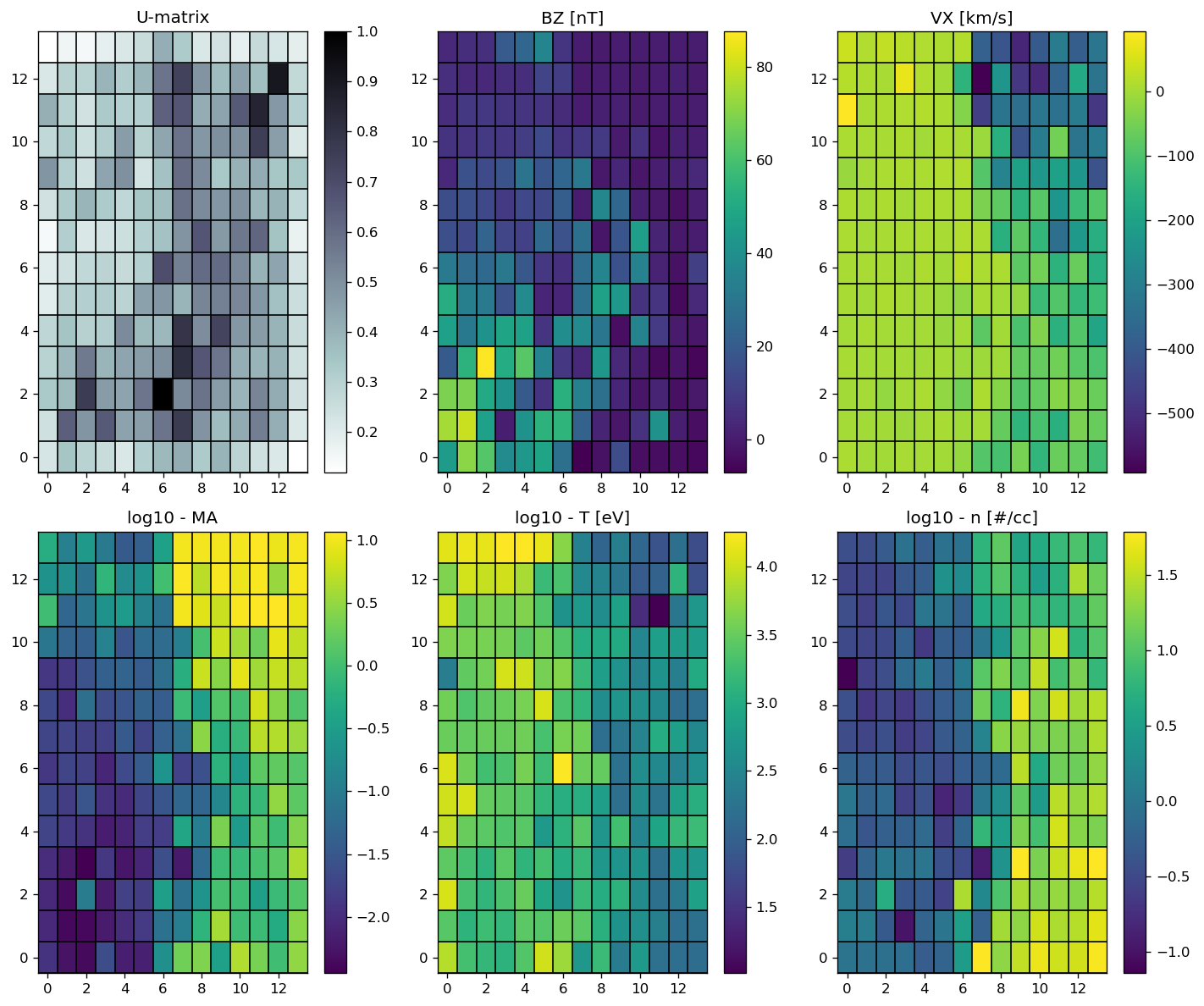}
\caption{ 2d heatmaps of the test data as seen through the SOM. In the U-matrix, plotted in the top left, nodes are coloured according to their distance to the nearest neighbours: the lighter nodes are more similar to the neighbours than darker nodes. A region of dark grey nodes partitions the U-Matrix into two areas of lighter color in the top left and bottom right. This means that there are two relatively homogeneous groups of nodes. To interpret what groups of data these nodes represent, we can look at the feature maps in the remaining plots. In these plots, the average feature value per node is depicted as a heatmap. It is apparent from the feature maps that the group of nodes on the left side of the U-Matrix correspond to regions of low density and high temperature. The nodes to the right correspond to moderate-to-high densities, low-to-moderate temperatures and negative values of VX. }
\label{som_feature_map}
\end{figure}

Another method uses the ordered nature of the SOM to create a heatmap of distances between the nodes. Since the nodes of a SOM have an ordered topological relationship, we can compute the average distance between a node and its immediate neighbors and create a heatmap of these average neighbor distances. The 2d matrix of these values is referred to as the U-Matrix. The U-Matrix for the test data can be seen in the top left of Figure \ref{som_feature_map}. Moreover, since each data point can be uniquely associated with its corresponding BMU in the SOM, then we can compute the average of all data per node. This average value per node can be used to create heatmaps of the SOM for any feature from the data, as seen in the remaining plots of Figure \ref{som_feature_map}.

\section{Clustering of SOM Nodes}

In wanting to apply direct clustering methods to our data, we had difficulties involving size, dimensionality, and multicollinearity. We resolved the latter two using PCA and have addressed the first by training a SOM to act as a further discretized representation of the data. With a SOM representation, we now can consider a much wider choice of methods to cluster the data as training size is no longer a constraining factor. Once a clustering method is trained, it can separate the SOM nodes automatically, predicting which nodes belongs to which cluster. These node predictions can then be propagated to the data that the nodes represent, i.e. if a node A is assigned to cluster 1, then all data for which node A is the BMU will be assigned to cluster 1. We will use a hierarchical form of clustering.


\subsection{Hierarchical Clustering}

Hierarchical clustering involves using one of two approaches. Agglomerative clustering assumes that all data points are individual clusters and that they can be iteratively merged based on the clusters' similarity. This is called the ``bottom-up" approach to hierarchical clustering. The complement to this is divisive clustering, which assumes all data initially belongs to a single cluster and iteratively separates data into heterogeneous subclusters. This is called the ``top-down" approach. We use the hierarchical agglomerative method as implemented in the scikit-learn package.

The quantification of similarity between a cluster A and B is computed using a linkage function, for which there are several common types: Maximum (or complete) linkage will define the distance from A to B to be the largest pairwise distance between a point in A and another in B. Minimum (or single) linkage defines the distance as the minimum pairwise distance. Average linkage will define the distance to be the average of the data of A to the average of the data of B. Ward's linkage does not use distance in the previous senses. Rather, it defines the distance between two clusters A and B as the variance of a new cluster obtained combining A and B. Since the variance is computed as the sum of square deviations from the mean (SSD), the clusters that will be merged are the ones which minimise this sum. We use a Ward linkage for our clustering model.



Because this method is hierarchical, one needs to define stopping criteria for cluster merging. This is done by visualizing the order of merging using a dendrogram where clusters are shown on the x axis as individual vertical lines and their merge order can be inferred from when their lines are horizontally merged. The position on the y axis of the merging is the SSD of the merged clusters. The dendrogram of the entire agglomerative merging process is visualized first and then a threshold distance is chosen so that only clusters with SSD below this cutoff will be considered.

A disadvantage of using hierarchical clustering is that getting predictions on data not seen during training can be difficult as the method is inherently transductive, i.e. it is trained on a specific dataset and does not generalize to unseen data. This limitation can sometimes be circumvented depending on the linkage used but certainly not in general. For example, in using a centroid linkage, one could simply assign new data to clusters whose centroids are closest, but this concept of closeness or similarity becomes vague in the context of other linkages, as some linkage types are capable of uncovering non-convex cluster distributions. To avoid this issue, we can use a hierarchical agglomerative clustering method to organize the nodes of the SOM into clusters and then propagate the cluster assignments of each node to the data that each node represents. In this way, data not seen during training is always assignable to some node of the SOM and all nodes belong to some cluster. By using a SOM to represent the data, we are able to use what is traditionally a transductive method in an inductive manner.


\subsection{Clustering Results}

\begin{figure}
\noindent\includegraphics[width=\textwidth]{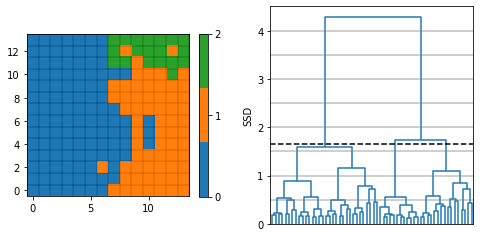}
\caption{
\textbf{Right}: A dendrogram of the clustered nodes using a Ward linkage. Separate clusters only up to the five most recent mergings are shown in order to not clutter the bottom of the plot. We choose a cutoff SSD of 1.65 to extract three clusters, as shown by the horizontal dashed black line. The number of times the line intersects with the vertical lines of clusters is the number of clusters recovered. The cluster assignments are visualized in the left image.
\textbf{Left}: Cluster assignments of the SOM nodes shown on the 2d node grid. The region of low density and high temperature observed in Figure \ref{som_feature_map} has been assigned to cluster 0 (blue), the region of low VX is largely cluster 2 (green) and the region of high density is largely cluster 1 (orange). The color scheme used to represent the different clusters will remain the same. }
\label{dendro_and_somclust}
\end{figure}

\begin{figure}
\noindent\includegraphics[width=\textwidth]{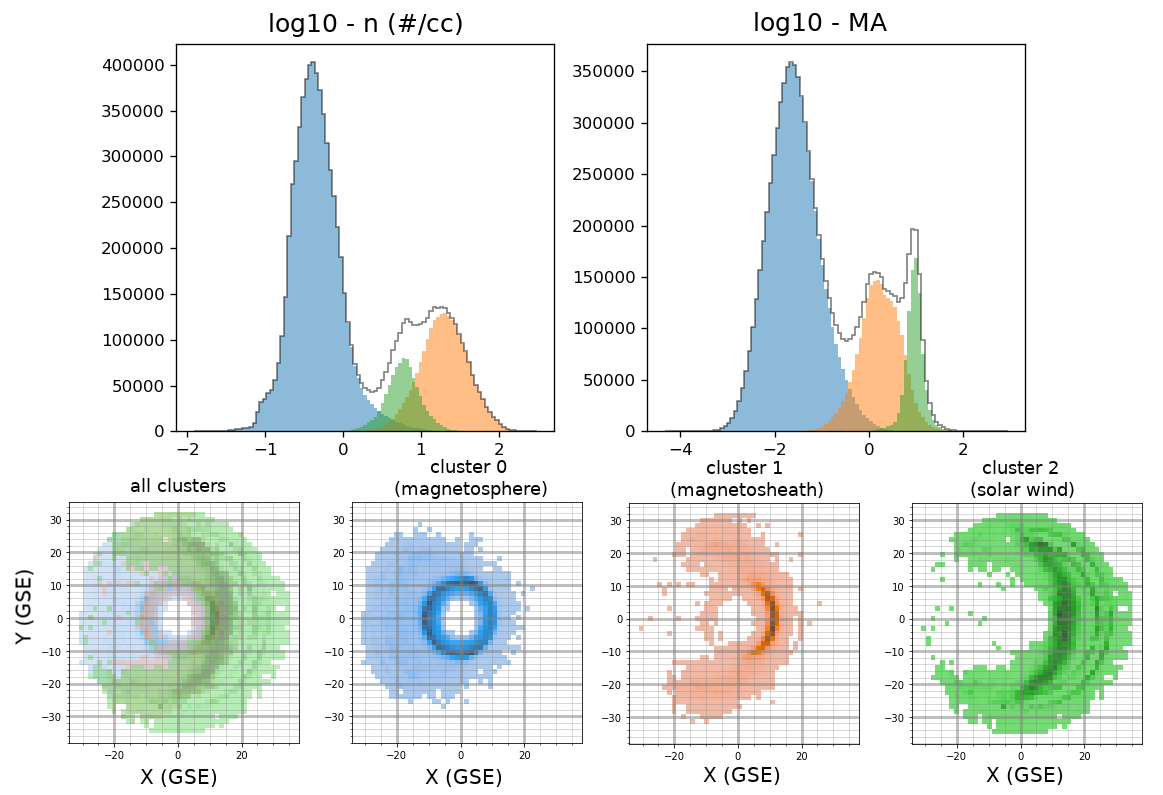}
\caption{
\textbf{Top / univariate histograms}: Histograms of the log10 density and log10 Alfv\'en Mach number. The histogram over the entire test set is in black and the histograms of the three clusters are represented in color. The magnetosphere is in blue (cluster 0), the magnetosheath is in orange (cluster 1) and the solar wind is in green (cluster 2).
\textbf{Bottom / bivariate histograms}: ($X_{GSE} [R_E]$, $Y_{GSE} [R_E]$) bivariate histograms of cluster occupancy where the sun is on the right. The leftmost plot shows the histogram over the entire test set and each other plot shows an occupancy histogram for a particular cluster. The cluster color scheme used is the same as in Figure \ref{dendro_and_somclust}. A darker shade of color indicates a higher count in the bivariate bin. }
\label{testpreds_histograms}
\end{figure}

\begin{figure}
\noindent\includegraphics[width=\textwidth]{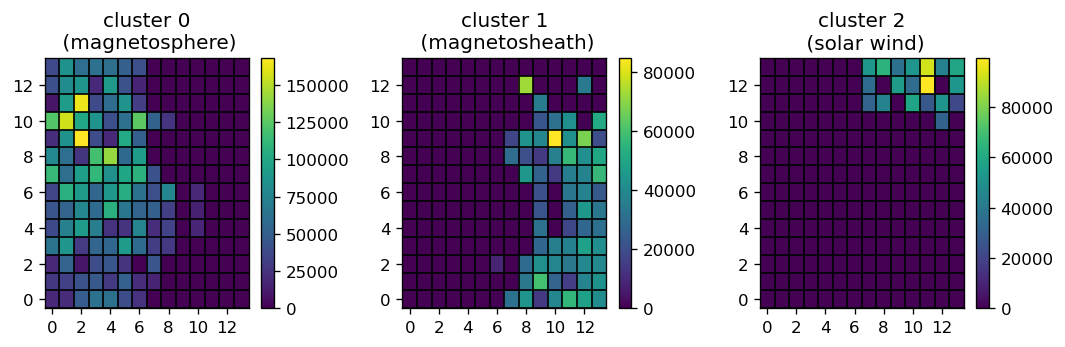}
\caption{ 
For each cluster, the number of points per node is shown. Note that the magnetosphere-classified nodes (10,6), (10,5), and (10,4) within the magnetosheath cluster contain few hits and the magnetosheath-classified node (6,2) within the magnetosphere cluster also contains few hits. However, the magnetosheath nodes (12,12) and (8,12) within the solar wind cluster are responsible for a sizable number of hits. }
\label{som_hits_test}
\end{figure}

\begin{figure}
\noindent\includegraphics[width=\textwidth]{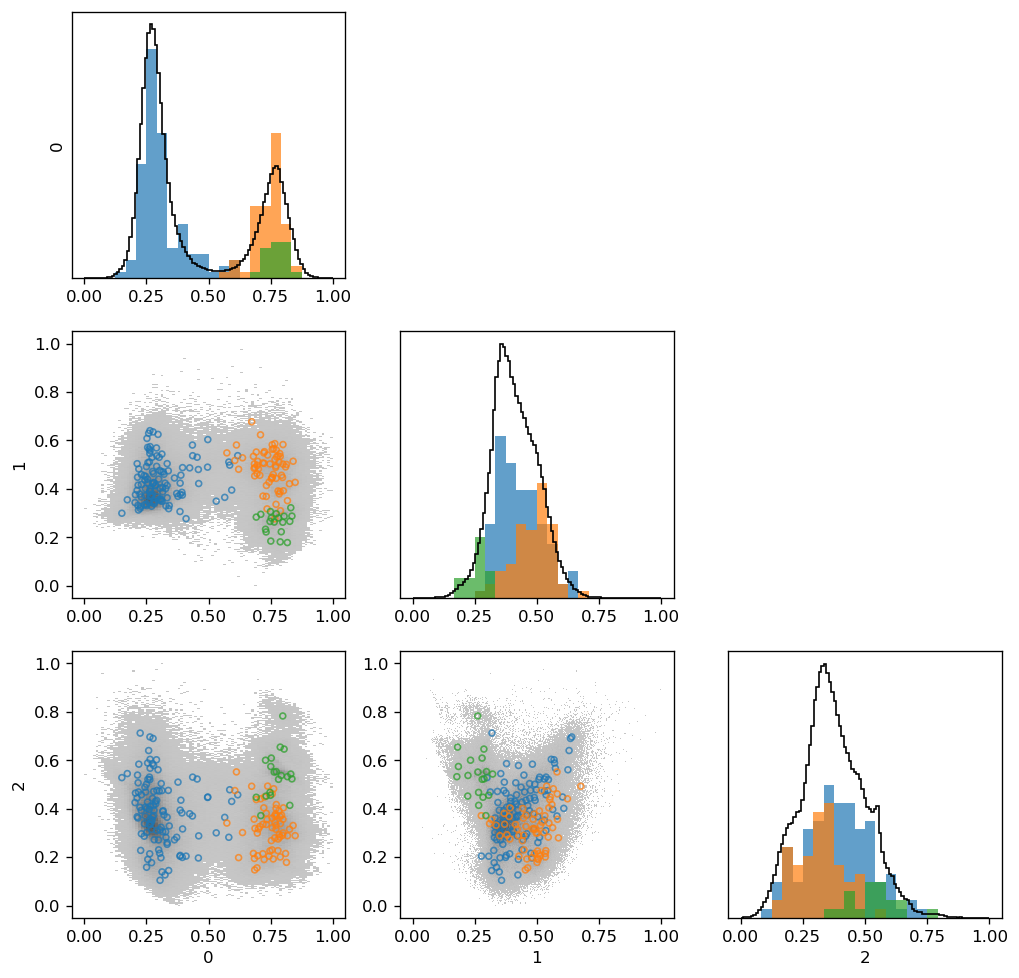}
\caption{ Cluster assignments of SOM nodes over the first three min-max normalized principal components of the test set. Comparing the plot of the (0,1) component-transformed data (center-left plot) to the biplot over the first two principal components in Figure \ref{pca_biplots}, we observe that the region on the left is the magnetosphere, the upper right is the magnetosheath, and the lower right is the solar wind. The marginal histograms of all clusters are shown along the diagonal using the same bin ratio (100 bins for data and 25 for SOM nodes) as in Figure \ref{som_and_data_pairplot}. }
\label{som_clust_and_pairplots}
\end{figure}

The dendrogram of the clustered SOM nodes and their cluster assignments are shown in Figure \ref{dendro_and_somclust}. From the dendrogram, we make cluster predictions using a distance threshold of 1.65 and propagate the cluster assignments of the SOM nodes to the test data. Histograms of the predictions for each cluster are shown in Figure \ref{testpreds_histograms}. These clusters were obtained in an unsupervised manner and aposteriori analysis shows that they correspond with specific regions, those being the magnetosphere, magnetosheath, and solar wind. Counts of the number of data points in the test set mapped per node is shown in figure \ref{som_hits_test}. The clustering of the SOM nodes in PCA space is shown in Figure \ref{som_clust_and_pairplots}. We previously made conjectures as to what portions of the biplot from Figure \ref{pca_biplots} are associated with the solar wind, magnetosheath, and magnetosphere, and they are confirmed with the clustering depicted. In both the (0,1) and (0,2) plots of Figure \ref{som_clust_and_pairplots}, the magnetosheath cluster has overlap with both the magnetosphere and the solar wind clusters but the magnetosphere and solar wind clusters have little overlap with each other, as one can expect from the physics of the magnetospheric system. Higher order components possess less variance and show considerable overlap as seen in the (1,2) plot. This is a consequence of using PCA for dimensionality reduction: The first PCA components will generally capture the majority of the variance and subsequent components will be less significant.

\begin{figure}
\noindent\includegraphics[width=\textwidth]{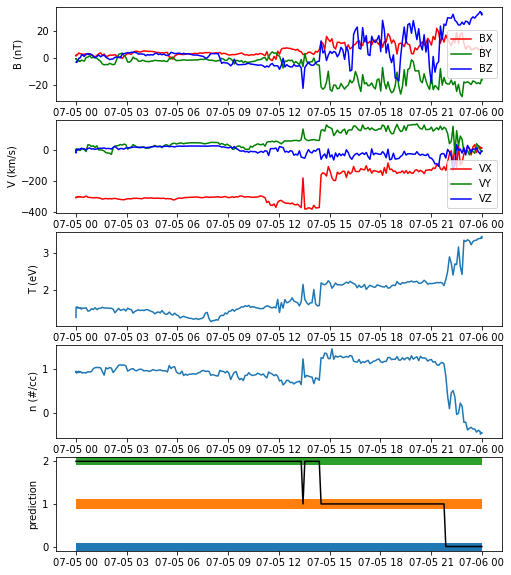}
\caption{ THEMIS-C measurements from 2008-07-05 to 2008-07-06. The temperature and density are in log10 scale. The predictions are shown in the bottom plot with the same cluster color scheme as Figure \ref{dendro_and_somclust}. The model successfully predicts the solar wind, magnetosphere, and magnetosheath measurements according to our visual verification. Noticeably, it also catches the ``blip" when THEMIS-C is briefly in the magnetosheath before again crossing the bow shock and going back into the magnetosheath at 14:00 UT. }
\label{thc_crossing}
\end{figure}

\begin{figure}
\noindent\includegraphics[width=\textwidth]{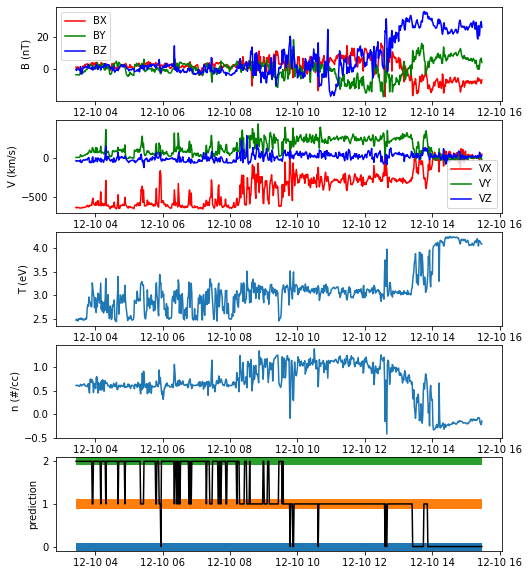}
\caption{ MMS 1 measurements from 2018-12-10 to 2018-12-11. The temperature and density are in log10 scale. The predictions are shown in the bottom plot with the same cluster color scheme as Figure \ref{dendro_and_somclust}. MMS 1 crosses the bow shock at about 8:00 UT and the magnetopause shortly after 13:00. The majority of the classifications prior to crossing the bow shock are solar wind, but there are a number of incorrect and spurious magnetosheath classifications as well as one magnetosphere classification. After 8:00 UT, the majority of classifications changes to magnetosheath with rarer solar wind and magnetosphere predictions occurring. In the interval when MMS 1 is in the magnetosheath, magnetosphere misclassificatons correspond with sudden drops in density measurements. }
\label{mms1_crossing}
\end{figure}

In GSE coordinates, the solar wind tends to be in the sunward (here, rightward) direction, the magnetosphere in the tailward (leftward) direction, and the magnetosheath is a curved transition region between the two. The histograms of log10 density and log10 Alfv\'en Mach number of Figure \ref{testpreds_histograms} reflect this and show the clustering is very effective in separating supersonic, moderate density plasma (solar wind) from shocked, dense plasma (magnetosheath) and very subsonic, thin plasma (magnetosphere). Note that since the Alfv\'en Mach number is plotted in log10 scale, the supersonic to subsonic transition occurs as a change in sign. Overlap between these distributions can certainly occur and this is reflected in their histograms. Incorrect predictions are also visible in Figure \ref{testpreds_histograms}, such as scattered magnetosheath and solar wind predictions occurring in the nightside at -20 $R_E \leq Y_{GSE} \leq$ 20 $R_E$, a swath of magnetosheath predictions at -10 $R_E \leq X_{GSE} \leq$ -5 $R_E$, and magnetosphere predictions well out into the dayside. In analyzing time series, these are generally spurious and rarely part of consecutive misclassifications. We show two sample predictions of time series, one for THEMIS-C where the prediction is exactly correct (Figure \ref{thc_crossing}) and one where the majority of classifications are correct but suffer from spurious misclassifications (Figure \ref{mms1_crossing}). Analyzing when MMS 1 is in the solar wind in Figure \ref{mms1_crossing}, it's apparent that the magnetosheath-misclassifications correspond to higher temperature and lower absolute value of the velocity, as in the magnetosheath. When MMS 1 is in the magnetosheath, the solar wind-misclassifications correspond to higher absolute value in velocity and the magnetosphere-misclassifications correspond to lower density, again consistent with the characteristics of the region to which the measurements are incorrectly assigned.

In Figure \ref{dendro_and_somclust}, it is evident that the different clusters are largely segregated spatially in the node grid but exceptions are present. There are multiple nodes that are at best somewhat adjacent to the remainder of their cluster. Notably, the magnetosheath cluster has nodes at grid positions (12,12) and (8,12) that are surrounded by the solar wind cluster. The magnetosheath cluster also has a node that is surrounded by the magnetosphere cluster at (6,2) and a vertical streak of magnetosphere-classified nodes starting at (10,4). Results like this are not entirely unexpected as we are analyzing observations and the magnetosheath acts as a transition region between the magnetosphere and solar wind. 




\subsection{Analyzing Particular Nodes}

We next analyze four instances of anomalous node positions in the SOM, namely the separated magnetosheath nodes at positions (12,12), (8,12), and (6,2) as well as the vertical streak of magnetosphere nodes at positions (10,6), (10,5), and (10,4).

\begin{figure}
\noindent\includegraphics[width=\textwidth]{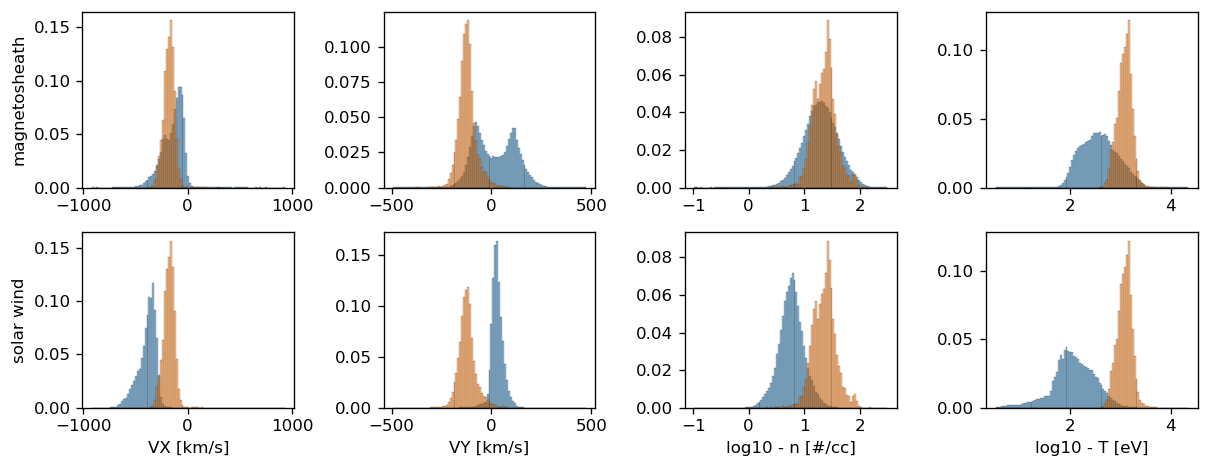}
\caption{ The VX, VY, log10 density, and log10 temperature empirical probability distributions of all magnetosheath-classified test data are plotted along the top row in blue. Similar features but for all solar wind-classified test data are plotted along the bottom row, also in blue. The empirical probability distribution of all test data that maps to node (12,12) is plotted in all plots as the orange distribution. The probability distributions are plotted here because of the large size differences between the number of magnetosheath observations (2.2 million) and solar wind observations (890k) and number of data mapping to node (12,12) (34k). }
\label{node1212}
\end{figure}

The node at (12,12) has the largest U-Matrix value seen in Figure \ref{som_feature_map}, indicating that it is farther from its neighbors than all other nodes in the SOM. This is not surprising since it is classified as a magnetosheath node and is surrounded by solar wind-classified nodes. There are about 2.2 million magnetosheath points in the test set and 34k (1.5\%) of them map to this node. Categorizing this node's data by spacecraft, we find that almost all are MMS observations with only about 100 belonging to THEMIS. We plot the empirical probability distributions of all magnetosheath and solar wind measurements in the test set in Figure \ref{node1212} as well as the data belonging to this node for comparison. From the figure, we can see that there is much more overlap with the distributions of node (12,12) with the magnetosheath observations than that of solar wind, indicating that although the node's position in the grid is unusual, it corresponds well with magnetosheath observations.

\begin{figure}
\noindent\includegraphics[width=\textwidth]{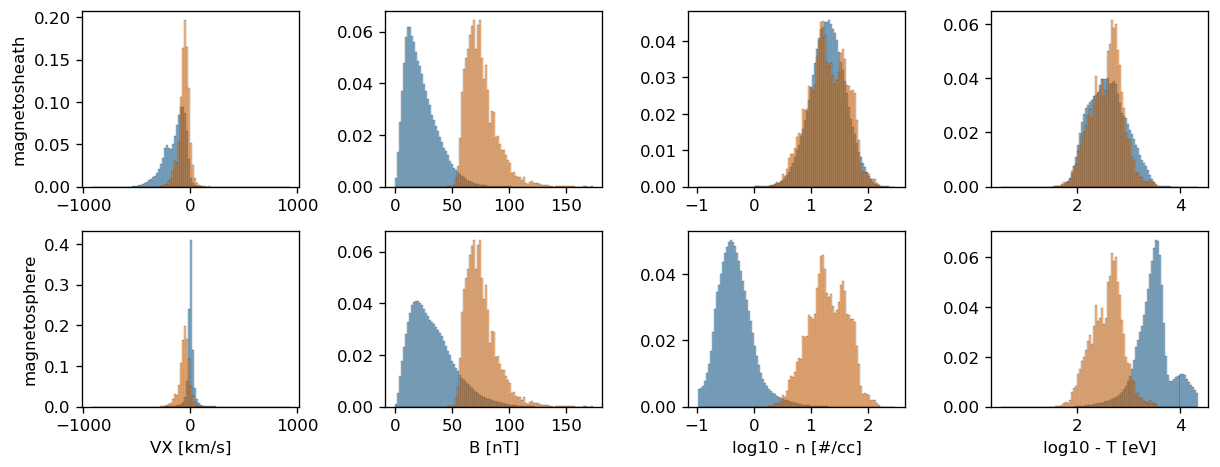}
\caption{ The VX, B, log10 density, and log10 temperature empirical probability distributions of all magnetosheath-classified test data are plotted along the top row in blue. Similar features but for all magnetosphere-classified test data are plotted along the bottom row, also in blue. The empirical probability distribution of the 8.5k magnetosheath-classified observations of node (6,2) are plotted in orange for each feature. The VX, log10 density, and log10 temperature distributions for this node all align more with the magnetosheath data than that classified as magnetosphere whereas the B distribution reflects high magnitude observations. Overall, this node has captured data with magnetosheath characteristics in velocity, density, and temperature, but also possessing high field magnitudes. }
\label{node62}
\end{figure}

Node (6,2) is another topologically isolated magnetosheath node that also possesses a very high U-Matrix value, except that this one is surrounded by magnetosphere-classified nodes. It is responsible for only about 8.5k (0.39\%) points of the magnetosheath-classified data of the test set and is almost evenly split by spacecraft with 56\% points belonging to THEMIS and 44\% to MMS. The empirical probability distributions of all magnetosheath-classified and magnetosphere-classified data in the test set are plotted alongside the observations mapped to this node in Figure \ref{node62} and multiple distinctions can immediately be made: data mapping to this node exhibit more magnetosheath characteristics in velocity, density, and temperature and also possess high magnetic field magnitudes. It seems correct that this node is classified as magnetosheath and the sparsity of points mapping to this node is understood in the context that magnetosheath observations possessing such large magnetic field magnitudes is relatively rare. The large U-Matrix value is justified with these observations.

\begin{figure}
\noindent\includegraphics[width=\textwidth]{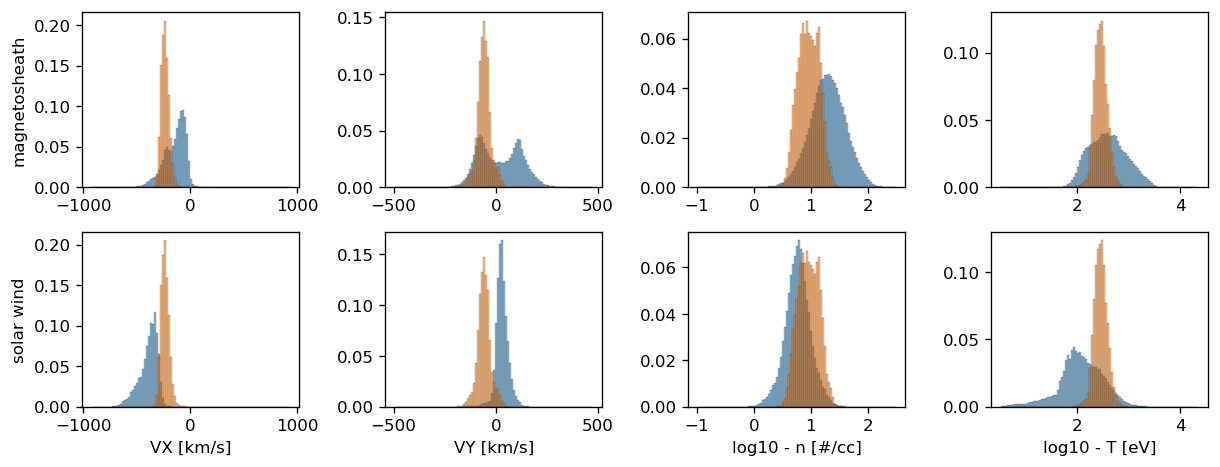}
\caption{ The VX, VY, log10 density and log10 temperature empirical probability distributions of all magnetosheath-classified data from the test set are plotted in blue along the top row. The solar wind-classified test data are plotted in blue along the bottom. The empirical probability distribution of the 69k magnetosheath-classified observations of node (8,12) are plotted in orange for each feature. The log10 density and log10 temperature distributions of the data from this node have sizeable mixing between both magnetosheath and solar wind observations whereas the VX and VY distributions are more distinctly magnetosheath than solar wind. }
\label{node812}
\end{figure}

Node (8,12) is diagonally topologically adjacent to the magnetosheath cluster but otherwise surround by solar wind nodes. This SOM uses a square topology, so this diagonal proximity does not factor into its U-Matrix value. It maps 69k (3.1\%) points from the magnetosheath-classified data of the test set with 11\% being THEMIS observations and 89\% being MMS. The VX, VY, log10 temperature, and log10 density empirical probability distributions of this data mapping to this node are shown in Figure \ref{node812} alongside all magnetosheath-classified and solar wind-classified test data. They indicate magnetosheath observations with respect to the VX and VY distributions, but the log10 temperature and log10 density distributions somewhat resemble a blend solar wind and magnetosheath. This lack of uniform agreement across these features can justify that node (8,12) is adjacent to solar wind-classified nodes but the VX and VY distributions in particular indicate that it is correct to classify it as a magnetosheath node.

\begin{figure}
\noindent\includegraphics[width=\textwidth]{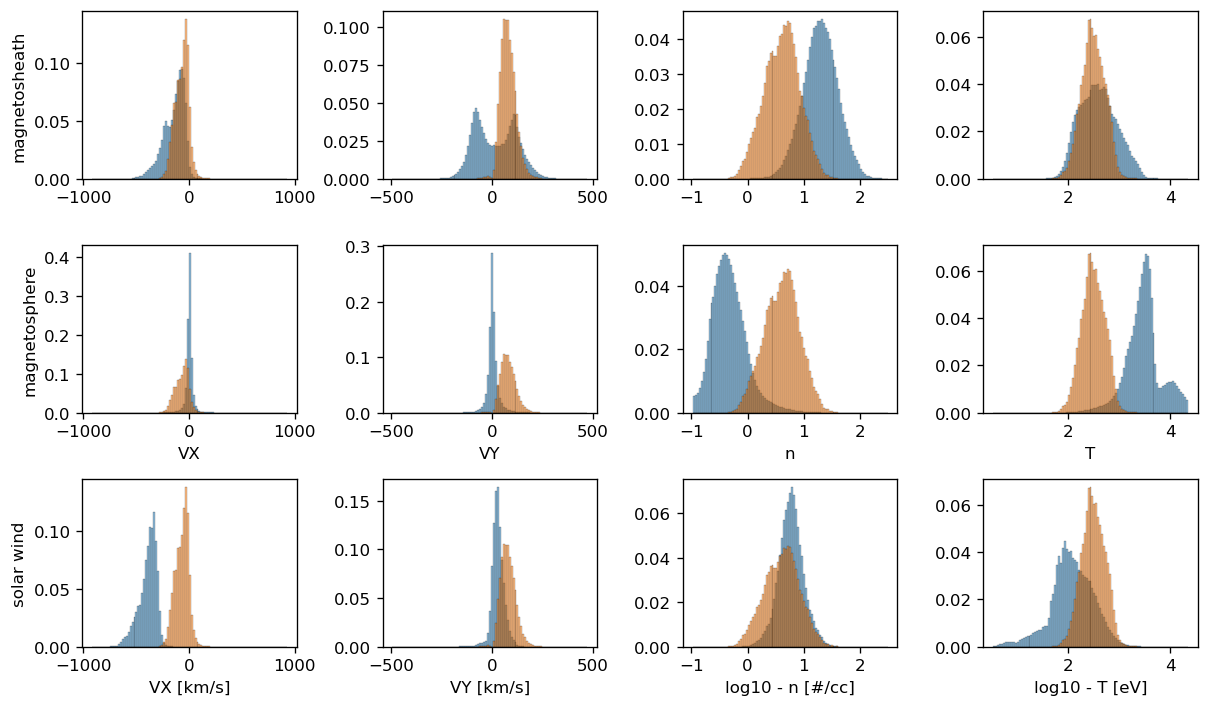}
\caption{ The VX, VY, log10 density and log10 temperature empirical probability distributions of all magnetosheath-, magnetosphere-, and solar wind-classified test data are plotted in blue along the top, middle, and bottom rows, respectively. All test data that map to nodes (10,6), (10,5) and (10,4) are collectively plotted here as the orange empirical probability distributions. These data are anomalous and exhibit characteristics found in all magnetosheath, magnetosphere, and solar wind observations. The VY, log10 density, and log10 temperature align well with the solar wind distributions, but the VX distribution is far too low. The VX, VY, and log10 temperature distributions correspond with magnetosheath observations, but there are very low densities. All of these distributions seem to have the least in common with the magnetosphere cluster, being along the extrema in all cases. }
\label{nodes10x}
\end{figure}

Lastly, we analyze the magnetosphere-classified nodes at positions (10,6), (10,5) and (10,4) that occur topologically within the magnetosheath cluster. Together, these nodes account for 48k (0.75\%) of the 6.4 million magnetosphere-classified points of the test set with 76\% being THEMIS observations and 24\% belonging to MMS. Their VX, VY, log10 density and log10 temperature empirical probability distributions are plotted in Figure \ref{nodes10x} along with the distributions of all three clusters in the test set. The data that map to these nodes are unusual in that the node distributions do not fully overlap with all of the distributions for any cluster. These data are classified as magnetosphere, but exist along the extrema of all the magnetosphere distributions shown. They resemble the VY, log10 density, and log10 temperature distributions of the solar wind, but the VX would be quite low for solar wind. The VX, VY, and log10 temperature distributions match up well with the magnetosheath distributions, but the log10 density is conspicuously low. Across all of the clusters, the measurements have much more in common with magnetosheath observations than magnetosphere or solar wind and are likely misclassifications. These nodes are responsible for 0.50\% of the total test set.

Overall, the magnetosheath cluster has nodes in several aberrant positions in the SOM grid in which they were surrounded by nodes belonging to other clusters. Investigating these nodes in detail, however, has shown that the data correspond well with magnetosheath observations and are deserving of being classified as such. It was also seen that three magnetosphere-classified nodes are likely misclassified and should be recognized as magnetosheath. These three nodes contain few points (50k points, or 0.50\% of the test set), together containing slightly more than the average number of points per node (49k), and so do not significantly impact the strength of the results. Furthermore, it should be noted that such a misclassification occurred between the magnetosheath and the magnetosphere and that the separation between solar wind and magnetosphere plasma is quite distinct in the map.


\subsection{Subpopulation Analysis}

\begin{figure}
\noindent\includegraphics[width=\textwidth]{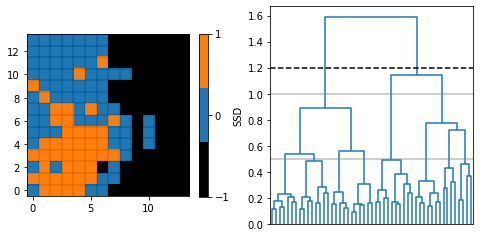}
\caption{ Like Figure \ref{dendro_and_somclust} but only focusing on the magnetosphere cluster of the test set.
\textbf{Right}: A dendrogram showing the merge order of the magnetosphere cluster. This tree is a subset of the dendrogram in Figure \ref{dendro_and_somclust}. We use a cutoff SSD of 1.2 and extract two clusters from the magnetosphere cluster.
\textbf{Left}: Subcluster assignments of the SOM nodes based on the distance chosen in the dendrogram. The nodes that did not belong to the magnetosphere cluster are masked out in black and assigned a label of -1. Looking back to the feature maps in Figure \ref{som_feature_map}, we can see that the blue cluster (0) is related to higher subsonic Alfv\'en Mach number and the orange cluster (1) is related to lower subsonic Alfv\'en Mach number.}
\label{dendro_and_somclust_magsphere}
\end{figure}

\begin{figure}
\noindent\includegraphics[width=\textwidth]{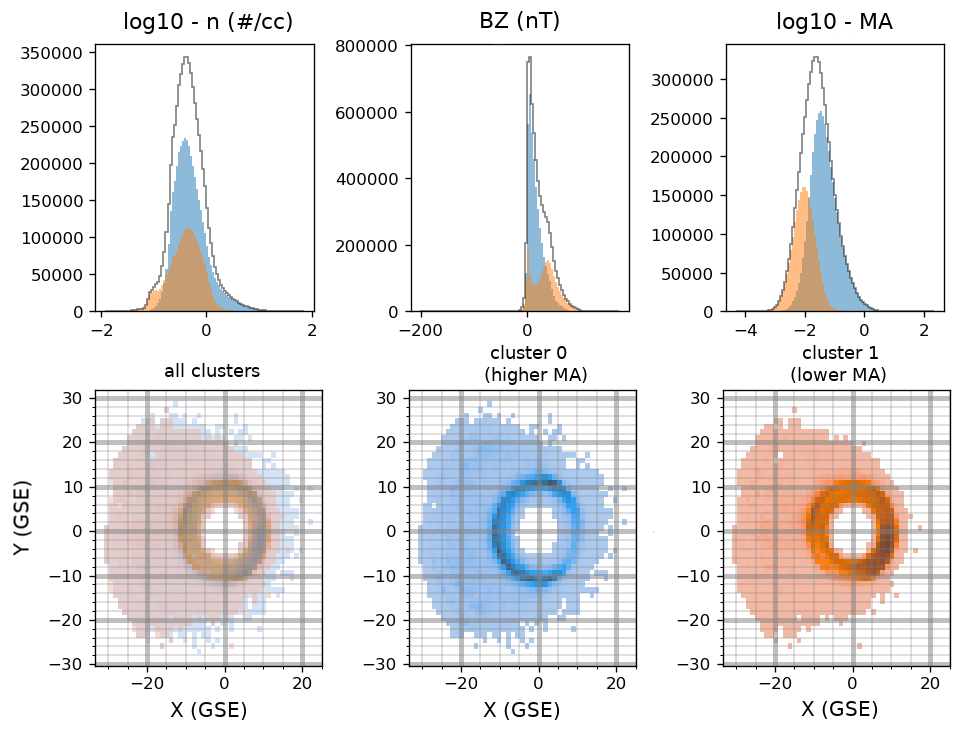}
\caption{ Like Figure \ref{testpreds_histograms}, but analyzing only the magnetosphere cluster of the test set.
\textbf{Bottom / bivariate histograms}: The occupancy of cluster 0 (blue) and 1 (orange) are plotted as bivariate histograms in ($X_{GSE}[R_E]$, $Y_{GSE}[R_E]$). They cover a similar region, but cluster 1 is much less pronounced on the dayside.
\textbf{Top / univariate histograms}: The histograms of the test data for log10 density, BZ, and log10 Alfv\'en Mach number are plotted in black and the cluster populations are plotted in their respective colors. As could be inferred from Figure \ref{som_feature_map}, cluster 0 is related to higher subsonic Alfv\'en Mach number and cluster 1 to lower subsonic values. }
\label{testpreds_histograms_magsphere}
\end{figure}

We also show in brief the capability of subpopulation analysis with this clustering method. Since we have used a hierarchical method to cluster the SOM nodes, we can pick any cluster and investigate the previously merged clusters that compose it. We ``unpack" the magnetosphere cluster in Figures \ref{dendro_and_somclust_magsphere} and \ref{testpreds_histograms_magsphere} to show how distinct magnetospheric populations were collectively recognized as the magnetosphere. From the histograms, we see that the feature that changes most clearly between the two clusters is the Alfv\'enic Mach number. Note that the subclusters of the magnetosphere in Figure \ref{dendro_and_somclust_magsphere} are not as evenly topologically separated like the original clustering solution seen in Figure \ref{dendro_and_somclust}. This is not surprising given the large overlap in features between these subclusters seen in the univariate histograms of Figure \ref{testpreds_histograms_magsphere} and indicates that the variance between these two subclusters is less than the variance between the magnetosphere, magnetosheath, and solar wind clusters. In simpler terms, it is easier to distinguish solar wind from magnetosheath and magnetosphere than it is to separate magnetospheric populations by Alfv\'en Mach number.

\section{Derived Boundary Crossings}

\begin{figure}
\noindent\includegraphics[width=\textwidth]{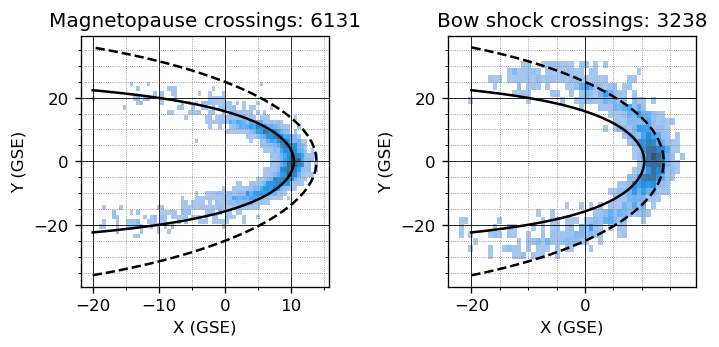}
\caption{ Bivariate histograms of the magnetopause (left) and bow shock (right) crossings in ($X_{GSE}[R_E]$, $Y_{GSE}[R_E]$). In both figures, the solid line is a Shue magnetopause with parameters n = 6.5 \#/cc, V = 390 km/s, and BZ = 0.057 nT and the dashed line is a simple fitted hyperbola of the form $\rho = 25 / (1 + 0.8 \cos \theta )$. The Shue parameters were chosen by taking the average of all solar wind-classified test set points. Many of the crossings are in line with expectations of magnetopause and bow shock position. }
\label{crossings}
\end{figure}

With a model that can predict when a measurement occurs in the magnetosphere, magnetosheath, or solar wind, we can study the time series of these predictions and infer when a spacecraft has crossed the magnetopause or bow shock. To account for the spurious misclassifications that can occur, we will select a crossing based on a cluster number time window of 30 minutes with a consistent classification of at least 90\%. For a crossing in which no misclassifications occur, there should be only a single change in cluster number for the interval, but for an interval containing misclassifications, the number of cluster changes will be more. If a crossing is identified in an interval, it is only kept if this number of cluster changes is 10\% of the interval size or less. That is, a bow shock crossing is extracted if the previous 15 minutes are labeled the same (being either solar wind or the magnetosheath) and if the following 15 minutes are consistently labeled as the other cluster, barring misclassifications that fall under this 10\% threshold. A similar technique is applied for the magnetopause crossings when considering prediction changes between the magnetosphere and magnetosheath clusters. A bivariate histogram of the ($X_{GSE}$, $Y_{GSE}$) positions of these crossings is depicted in Figure \ref{crossings} alongside a Shue magnetopause \cite{shue_mp} and a simple hyperbolic model of the bow shock and show good agreement with respect to both.

\begin{figure}
\noindent\includegraphics[width=\textwidth]{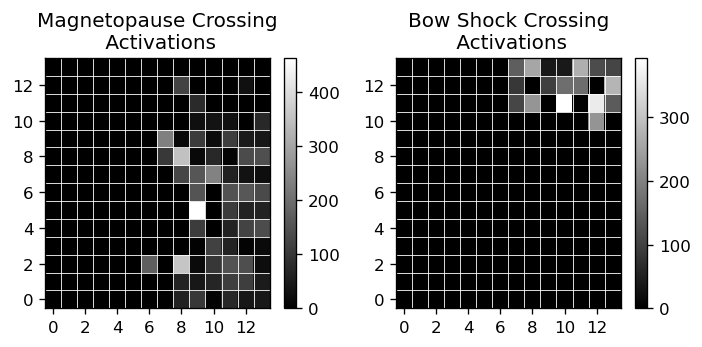}
\caption{ For each magnetopause (bow shock) crossing, we select the most recent magnetosheath (solar wind) point. Each point maps to, or ``activates'', some node in the SOM. The distribution of these counts is shown for the magnetosheath points for the magnetopause on the left and the solar wind points for the bow shock on the right. For the magnetosheath points, the most activated nodes are at positions (9,5), (8,8), and (8,2) and are together responsible for 1,170 crossings. For the solar wind points, the most activated nodes are at positions (10,11) and (12,11) and are responsible for 755 crossings.   }
\label{crossing_activations}
\end{figure}

For the bow shock crossings, we select the most recent solar wind point relative to the time of crossing and see how they're distributed in the SOM grid in Figure \ref{crossing_activations}. When cross-comparing these with the number of counts in the test set from Figure \ref{som_hits_test}, we see that the two most activated nodes of bow shock crossings are nodes (10,11) and (12,11). These nodes are responsible for 23\% of the crossings but only 12\% of the solar wind predictions: In the case of operational use of this model, a solar wind measurement assigned to one of these nodes could be flagged as having an increasing probability of resulting into a magnetopause crossing. Additionally, the node with the highest count in the test set for solar wind points, node (11,12), has only a small number of bow shock crossing points (5.2\%) relative to the previous nodes.

We perform a similar analysis for the magnetosheath points relative to the magnetopause crossings. The nodes with the highest number of counts of magnetosheath points associated with magnetopause crossings are the nodes (9,5), (8,8), and (8,2). These are responsible for 19\% of the magnetopause crossings but only 3\% of the magnetosheath predictions. The node with the largest number of magnetosheath points in the test set, node (10,9) at 3.6\%, only contains 21 magnetosheath points of the crossings, or 0.34\% of the magnetopause crossings. These three nodes could be used to provide a warning for identifying magnetopause crossings.


\section{Conclusion and Future Work}

We have taken magnetic field, ion velocity, ion density, and ion temperature measurements from THEMIS and MMS observations at different time resolutions in order to predict whether they occurred in the magnetosphere, magnetosheath, or the solar wind. The changing time step of THEMIS observations precludes the use of methods that need consistent timing and requires that we utilize tools that consider only the joint set of measurements. We have used a combination of unsupervised methods to cluster these data in a time independent way, and it works remarkably well with measurements from different spacecraft.


Seeking to express some of the nonlinear variances of the data linearly, we constructed additional features of magnetic field strength, ion speed, and the components of ion momentum density. Due to a range of orders of magnitude, features related to the ion density or temperature were converted to log10 scale, or the log10-absolute scale in the case of the ion momentum density. Partitioning the data into training and testing sets, these data were then min-max normalized based on the training set. PCA was used to yield a smaller set of uncorrelated features, both reducing the dimensionality and multicollinearity of the training data. Translating the loadings of a PCA transformation into a plot, we correctly predicted which regions of the PCA-transformed data would approximately correspond to certain anticipated clusters: higher temperatures (magnetosphere), higher densities (magnetosheath), and higher speeds (solar wind).

To find a SOM that best represents outliers in the training data, we used KMeans to build a micro training set of 10k points from the training data and made the remainder of the training set a macro training set, or validation set. 500 maps were trained on this micro training set and the hyperparameters of the SOM that returned an optimal value on the macro training set were retained. A SOM with these hyperparameters was then trained on the macro training set. Using the nodes of this map as representative samples of our training set, the number of clustering options available for use was expanded such that even transductive clustering methods could be utilized. The bulk of SOM nodes mapped to the locations of expected clusters from the PCA transform along the $0^{th}$ and $1^{st}$ components with the remainder of nodes distributed between them.

Next, we used hierarchical agglomerative clustering to cluster the nodes and make predictions on test data by propagating node cluster assignments to the data that the nodes represent. We used this method with a Ward linkage, which focuses on building clusters based on minimization of intra-cluster variance. The solar wind and magnetosphere clusters were well separated from each other whereas the magnetosheath cluster had multiple nodes that overlapped into both the solar wind and magnetosphere clusters, which is not surprising as the magnetosheath acts as a transition region. We investigated the magnetosheath nodes that were surrounded by either solar wind or magnetosphere nodes to ascertain if there were possible misclassifications, but detailed analysis showed that it was indeed correct that these nodes were classified as magnetosheath. The data mapping to these nodes sometimes showed a mix of characteristics between magnetosheath and the surrounding cluster, reaffirming that these magnetosheath nodes appear adjacent to the nodes of other clusters. We also investigated where three magnetosphere-classified nodes were situated within the magnetosheath cluster and inspection of these nodes' data revealed this classification to likely be incorrect. This data only represented 0.50\% of the test set and did not have sufficient alignment across all distributions for magnetosphere, magnetosheath, or solar wind, so any classification of this data was done without strong confidence. Being a hierarchical method, we also demonstrated how subcluster analysis is possible, an advantage not available to most other clustering methods. However, this does not mean that the uncovered subclusters will be as topologically ``smooth'' as parent clusters.

The validation of the model is done by visual inspection of both time series and histograms as the data is unlabeled. On the whole, the model seems to be generally accurate but is capable of spurious and largely non-consecutive misclassifications. Having separated data into magnetosphere, magnetosheath, or solar wind regions, we extract magnetopause and bow shock boundary crossings from the predictions on the full dataset. Having to account for both misclassifications and changing time resolution, we use a 30 minute window with a threshold that allows for very few misclassifications to occur in a crossing. We extracted 6,131 magnetopause crossings and 3,238 bow shock crossings. Analyzing the most recent solar wind points from the bow shock crossings in the SOM, we found that these points were distributed across most of the solar wind-classified SOM nodes with the two largest containing 23\% of the crossings. Performing the same analysis on the magnetosheath points in the context of magnetopause crossings, we found them less evenly distributed than what we saw for the solar wind nodes and noted that the three nodes with the highest number of counts for these points accounted for 19\% of the crossings but only 3\% of the magnetosheath predictions. In both cases, the nodes mapping the highest fraction of crossings can be useful in that data mapping to these nodes can be flagged as most likely to be related to a boundary crossing.

Future work will involve using our collection of bow shock crossings to create a new bow shock model with focus on the IMF clock angle. Additionally, we can investigate the subclusters composing the magnetosheath and solar wind uncovered by our model in greater detail.

\section{Open Research}

The authors thank both the THEMIS and MMS instrument teams for the instrument development and availability of the datasets. The THEMIS data was downloaded and processed using the PySPEDAS python library \cite{pyspedas}. The MMS data was downloaded and processed using the PyMMS python package \cite{pymms}.

In the interest of ease of access, we have made a python package GMClustering that will easily make predictions and is pip-installable directly from the repository. It is available at https://github.com/jae1018/GMClustering and includes both an example python driver file and a small Jupyter notebook to showcase its use. We also include our data for MMS1 for the year 2017 as an example dataset. Our modeling used various numerically-oriented python packages and we include the versions of those most relevant below.

\begin{itemize}
    \item Numpy \cite{numpy} : 1.24.3
    \item Scikit-Learn \cite{scikit} : 1.3.0
    \item XPySom \cite{xpysom} : 1.0.7
    \item Pandas \cite{pandas} : 2.0.3
    \item SciPy \cite{scipy} : 1.11.1
\end{itemize}







\acknowledgments

The authors state that this research was conducted in absence of any commercial or financial relationships that could be viewed as a conflict of interest.

Maria Elena Innocenti acknowledges support from the Deutsche Forschungsgemeinchaft (German Science Foundation, DFG) within the Collaborative Research Center SFB1491 and within the DFG project 497938371. Work at UNH was supported through AFOSR grant FA9550-18-1-0483 and from the NASA/THEMIS mission through subcontract SA405826326 from UC Berkeley.


\appendix


%
%


%
%



\bibliography{agusample.bib}

\begin{thebibliography}{}

\bibitem [\protect \citeauthoryear {%
Akiba%
, Sano%
, Yanase%
, Ohta%
\BCBL {}\ \BBA {} Koyama%
}{%
Akiba%
\ \protect \BOthers {.}}{%
{\protect \APACyear {2019}}%
}]{%
optuna}
\APACinsertmetastar {%
optuna}%
\begin{APACrefauthors}%
Akiba, T.%
, Sano, S.%
, Yanase, T.%
, Ohta, T.%
\BCBL {}\ \BBA {} Koyama, M.%
\end{APACrefauthors}%
\unskip\
\newblock
\APACrefYearMonthDay{2019}{}{}.
\newblock
{\BBOQ}\APACrefatitle {Optuna: A Next-generation Hyperparameter Optimization Framework} {Optuna: A next-generation hyperparameter optimization framework}.{\BBCQ}
\newblock
\BIn{} \APACrefbtitle {Proceedings of the 25th {ACM} {SIGKDD} International Conference on Knowledge Discovery and Data Mining.} {Proceedings of the 25th {ACM} {SIGKDD} international conference on knowledge discovery and data mining.}
\PrintBackRefs{\CurrentBib}

\bibitem [\protect \citeauthoryear {%
Amaya%
, Dupuis%
, Innocenti%
\BCBL {}\ \BBA {} Lapenta%
}{%
Amaya%
\ \protect \BOthers {.}}{%
{\protect \APACyear {2020}}%
}]{%
SW_AEandGSOM}
\APACinsertmetastar {%
SW_AEandGSOM}%
\begin{APACrefauthors}%
Amaya, J.%
, Dupuis, R.%
, Innocenti, M\BPBI E.%
\BCBL {}\ \BBA {} Lapenta, G.%
\end{APACrefauthors}%
\unskip\
\newblock
\APACrefYearMonthDay{2020}{}{}.
\newblock
{\BBOQ}\APACrefatitle {Visualizing and Interpreting Unsupervised Solar Wind Classifications} {Visualizing and interpreting unsupervised solar wind classifications}.{\BBCQ}
\newblock
\APACjournalVolNumPages{Frontiers in Astronomy and Space Sciences}{7}{}{}.
\newblock
\begin{APACrefURL} \url{https://www.frontiersin.org/articles/10.3389/fspas.2020.553207} \end{APACrefURL}
\newblock
\begin{APACrefDOI} \doi{10.3389/fspas.2020.553207} \end{APACrefDOI}
\PrintBackRefs{\CurrentBib}

\bibitem [\protect \citeauthoryear {%
Angelopoulos%
}{%
Angelopoulos%
}{%
{\protect \APACyear {2008}}%
}]{%
Angelopoulos_THEMIS}
\APACinsertmetastar {%
Angelopoulos_THEMIS}%
\begin{APACrefauthors}%
Angelopoulos, V.%
\end{APACrefauthors}%
\unskip\
\newblock
\APACrefYearMonthDay{2008}{Apr}{22}.
\newblock
{\BBOQ}\APACrefatitle {The THEMIS Mission} {The themis mission}.{\BBCQ}
\newblock
\APACjournalVolNumPages{Space Science Reviews}{141}{1}{5}.
\newblock
\begin{APACrefURL} \url{https://doi.org/10.1007/s11214-008-9336-1} \end{APACrefURL}
\newblock
\begin{APACrefDOI} \doi{10.1007/s11214-008-9336-1} \end{APACrefDOI}
\PrintBackRefs{\CurrentBib}

\bibitem [\protect \citeauthoryear {%
Angelopoulos%
}{%
Angelopoulos%
}{%
{\protect \APACyear {2014}}%
}]{%
Angelopoulos_ARTEMIS}
\APACinsertmetastar {%
Angelopoulos_ARTEMIS}%
\begin{APACrefauthors}%
Angelopoulos, V.%
\end{APACrefauthors}%
\unskip\
\newblock
\APACrefYearMonthDay{2014}{}{}.
\newblock
{\BBOQ}\APACrefatitle {The ARTEMIS Mission} {The artemis mission}.{\BBCQ}
\newblock
\BIn{} C.~Russell\ \BBA {} V.~Angelopoulos\ (\BEDS), \APACrefbtitle {The ARTEMIS Mission} {The artemis mission}\ (\BPGS\ 3--25).
\newblock
\APACaddressPublisher{New York, NY}{Springer New York}.
\newblock
\begin{APACrefURL} \url{https://doi.org/10.1007/978-1-4614-9554-3_2} \end{APACrefURL}
\newblock
\begin{APACrefDOI} \doi{10.1007/978-1-4614-9554-3_2} \end{APACrefDOI}
\PrintBackRefs{\CurrentBib}

\bibitem [\protect \citeauthoryear {%
M.~Argall%
, colinrsmall%
\BCBL {}\ \BBA {} Petrik%
}{%
M.~Argall%
\ \protect \BOthers {.}}{%
{\protect \APACyear {2022}}%
}]{%
pymms}
\APACinsertmetastar {%
pymms}%
\begin{APACrefauthors}%
Argall, M.%
, colinrsmall%
\BCBL {}\ \BBA {} Petrik, M.%
\end{APACrefauthors}%
\unskip\
\newblock
\APACrefYearMonthDay{2022}{{\APACmonth{05}}}{}.
\newblock
\APACrefbtitle {argallmr/pymms: v0.4.6 (2022-05-19).} {argallmr/pymms: v0.4.6 (2022-05-19).}
\newblock
\APACaddressPublisher{}{Zenodo}.
\newblock
\begin{APACrefURL} \url{https://doi.org/10.5281/zenodo.6564714} \end{APACrefURL}
\newblock
\begin{APACrefDOI} \doi{10.5281/zenodo.6564714} \end{APACrefDOI}
\PrintBackRefs{\CurrentBib}

\bibitem [\protect \citeauthoryear {%
M\BPBI R.~Argall%
\ \protect \BOthers {.}}{%
M\BPBI R.~Argall%
\ \protect \BOthers {.}}{%
{\protect \APACyear {2020}}%
}]{%
argall_mms_sitl_ml}
\APACinsertmetastar {%
argall_mms_sitl_ml}%
\begin{APACrefauthors}%
Argall, M\BPBI R.%
, Small, C\BPBI R.%
, Piatt, S.%
, Breen, L.%
, Petrik, M.%
, Kokkonen, K.%
\BDBL {}Burch, J\BPBI L.%
\end{APACrefauthors}%
\unskip\
\newblock
\APACrefYearMonthDay{2020}{}{}.
\newblock
{\BBOQ}\APACrefatitle {MMS SITL Ground Loop: Automating the Burst Data Selection Process} {Mms sitl ground loop: Automating the burst data selection process}.{\BBCQ}
\newblock
\APACjournalVolNumPages{Frontiers in Astronomy and Space Sciences}{7}{}{}.
\newblock
\begin{APACrefURL} \url{https://www.frontiersin.org/articles/10.3389/fspas.2020.00054} \end{APACrefURL}
\newblock
\begin{APACrefDOI} \doi{10.3389/fspas.2020.00054} \end{APACrefDOI}
\PrintBackRefs{\CurrentBib}

\bibitem [\protect \citeauthoryear {%
Auster%
\ \protect \BOthers {.}}{%
Auster%
\ \protect \BOthers {.}}{%
{\protect \APACyear {2008}}%
}]{%
THEMIS_FGM}
\APACinsertmetastar {%
THEMIS_FGM}%
\begin{APACrefauthors}%
Auster, H\BPBI U.%
, Glassmeier, K\BPBI H.%
, Magnes, W.%
, Aydogar, O.%
, Baumjohann, W.%
, Constantinescu, D.%
\BDBL {}Wiedemann, M.%
\end{APACrefauthors}%
\unskip\
\newblock
\APACrefYearMonthDay{2008}{Dec}{01}.
\newblock
{\BBOQ}\APACrefatitle {The THEMIS Fluxgate Magnetometer} {The themis fluxgate magnetometer}.{\BBCQ}
\newblock
\APACjournalVolNumPages{Space Science Reviews}{141}{1}{235-264}.
\newblock
\begin{APACrefURL} \url{https://doi.org/10.1007/s11214-008-9365-9} \end{APACrefURL}
\newblock
\begin{APACrefDOI} \doi{10.1007/s11214-008-9365-9} \end{APACrefDOI}
\PrintBackRefs{\CurrentBib}

\bibitem [\protect \citeauthoryear {%
Breuillard%
\ \protect \BOthers {.}}{%
Breuillard%
\ \protect \BOthers {.}}{%
{\protect \APACyear {2020}}%
}]{%
mms_region_classification}
\APACinsertmetastar {%
mms_region_classification}%
\begin{APACrefauthors}%
Breuillard, H.%
, Dupuis, R.%
, Retino, A.%
, Le~Contel, O.%
, Amaya, J.%
\BCBL {}\ \BBA {} Lapenta, G.%
\end{APACrefauthors}%
\unskip\
\newblock
\APACrefYearMonthDay{2020}{}{}.
\newblock
{\BBOQ}\APACrefatitle {Automatic Classification of Plasma Regions in Near-Earth Space With Supervised Machine Learning: Application to Magnetospheric Multi Scale 2016–2019 Observations} {Automatic classification of plasma regions in near-earth space with supervised machine learning: Application to magnetospheric multi scale 2016–2019 observations}.{\BBCQ}
\newblock
\APACjournalVolNumPages{Frontiers in Astronomy and Space Sciences}{7}{}{}.
\newblock
\begin{APACrefURL} \url{https://www.frontiersin.org/articles/10.3389/fspas.2020.00055} \end{APACrefURL}
\newblock
\begin{APACrefDOI} \doi{10.3389/fspas.2020.00055} \end{APACrefDOI}
\PrintBackRefs{\CurrentBib}

\bibitem [\protect \citeauthoryear {%
Burch%
, Moore%
, Torbert%
\BCBL {}\ \BBA {} Giles%
}{%
Burch%
\ \protect \BOthers {.}}{%
{\protect \APACyear {2016}}%
}]{%
MMS}
\APACinsertmetastar {%
MMS}%
\begin{APACrefauthors}%
Burch, J\BPBI L.%
, Moore, T\BPBI E.%
, Torbert, R\BPBI B.%
\BCBL {}\ \BBA {} Giles, B\BPBI L.%
\end{APACrefauthors}%
\unskip\
\newblock
\APACrefYearMonthDay{2016}{Mar}{01}.
\newblock
{\BBOQ}\APACrefatitle {Magnetospheric Multiscale Overview and Science Objectives} {Magnetospheric multiscale overview and science objectives}.{\BBCQ}
\newblock
\APACjournalVolNumPages{Space Science Reviews}{199}{1}{5-21}.
\newblock
\begin{APACrefURL} \url{https://doi.org/10.1007/s11214-015-0164-9} \end{APACrefURL}
\newblock
\begin{APACrefDOI} \doi{10.1007/s11214-015-0164-9} \end{APACrefDOI}
\PrintBackRefs{\CurrentBib}

\bibitem [\protect \citeauthoryear {%
Camporeale%
, Carè%
\BCBL {}\ \BBA {} Borovsky%
}{%
Camporeale%
\ \protect \BOthers {.}}{%
{\protect \APACyear {2017}}%
}]{%
sw_categories_supervised_ml}
\APACinsertmetastar {%
sw_categories_supervised_ml}%
\begin{APACrefauthors}%
Camporeale, E.%
, Carè, A.%
\BCBL {}\ \BBA {} Borovsky, J\BPBI E.%
\end{APACrefauthors}%
\unskip\
\newblock
\APACrefYearMonthDay{2017}{}{}.
\newblock
{\BBOQ}\APACrefatitle {Classification of Solar Wind With Machine Learning} {Classification of solar wind with machine learning}.{\BBCQ}
\newblock
\APACjournalVolNumPages{Journal of Geophysical Research: Space Physics}{122}{11}{10,910-10,920}.
\newblock
\begin{APACrefURL} \url{https://agupubs.onlinelibrary.wiley.com/doi/abs/10.1002/2017JA024383} \end{APACrefURL}
\newblock
\begin{APACrefDOI} \doi{https://doi.org/10.1002/2017JA024383} \end{APACrefDOI}
\PrintBackRefs{\CurrentBib}

\bibitem [\protect \citeauthoryear {%
Cranmer%
\ \BBA {} Winebarger%
}{%
Cranmer%
\ \BBA {} Winebarger%
}{%
{\protect \APACyear {2019}}%
}]{%
solar_corona}
\APACinsertmetastar {%
solar_corona}%
\begin{APACrefauthors}%
Cranmer, S\BPBI R.%
\BCBT {}\ \BBA {} Winebarger, A\BPBI R.%
\end{APACrefauthors}%
\unskip\
\newblock
\APACrefYearMonthDay{2019}{}{}.
\newblock
{\BBOQ}\APACrefatitle {The Properties of the Solar Corona and Its Connection to the Solar Wind} {The properties of the solar corona and its connection to the solar wind}.{\BBCQ}
\newblock
\APACjournalVolNumPages{Annual Review of Astronomy and Astrophysics}{57}{1}{157-187}.
\newblock
\begin{APACrefURL} \url{https://doi.org/10.1146/annurev-astro-091918-104416} \end{APACrefURL}
\newblock
\begin{APACrefDOI} \doi{10.1146/annurev-astro-091918-104416} \end{APACrefDOI}
\PrintBackRefs{\CurrentBib}

\bibitem [\protect \citeauthoryear {%
{de Bodt}%
, Cottrell%
, Letremy%
\BCBL {}\ \BBA {} Verleysen%
}{%
{de Bodt}%
\ \protect \BOthers {.}}{%
{\protect \APACyear {2004}}%
}]{%
som_and_vq}
\APACinsertmetastar {%
som_and_vq}%
\begin{APACrefauthors}%
{de Bodt}, E.%
, Cottrell, M.%
, Letremy, P.%
\BCBL {}\ \BBA {} Verleysen, M.%
\end{APACrefauthors}%
\unskip\
\newblock
\APACrefYearMonthDay{2004}{}{}.
\newblock
{\BBOQ}\APACrefatitle {On the use of self-organizing maps to accelerate vector quantization} {On the use of self-organizing maps to accelerate vector quantization}.{\BBCQ}
\newblock
\APACjournalVolNumPages{Neurocomputing}{56}{}{187-203}.
\newblock
\begin{APACrefURL} \url{https://www.sciencedirect.com/science/article/pii/S092523120300479X} \end{APACrefURL}
\newblock
\begin{APACrefDOI} \doi{https://doi.org/10.1016/j.neucom.2003.09.009} \end{APACrefDOI}
\PrintBackRefs{\CurrentBib}

\bibitem [\protect \citeauthoryear {%
Fairfield%
\ \protect \BOthers {.}}{%
Fairfield%
\ \protect \BOthers {.}}{%
{\protect \APACyear {2001}}%
}]{%
Fairfield_LowMachNumberShocks}
\APACinsertmetastar {%
Fairfield_LowMachNumberShocks}%
\begin{APACrefauthors}%
Fairfield, D\BPBI H.%
, Iver, H\BPBI C.%
, Desch, M\BPBI D.%
, Szabo, A.%
, Lazarus, A\BPBI J.%
\BCBL {}\ \BBA {} Aellig, M\BPBI R.%
\end{APACrefauthors}%
\unskip\
\newblock
\APACrefYearMonthDay{2001}{}{}.
\newblock
{\BBOQ}\APACrefatitle {The location of low Mach number bow shocks at Earth} {The location of low mach number bow shocks at earth}.{\BBCQ}
\newblock
\APACjournalVolNumPages{Journal of Geophysical Research: Space Physics}{106}{A11}{25361-25376}.
\newblock
\begin{APACrefURL} \url{https://agupubs.onlinelibrary.wiley.com/doi/abs/10.1029/2000JA000252} \end{APACrefURL}
\newblock
\begin{APACrefDOI} \doi{https://doi.org/10.1029/2000JA000252} \end{APACrefDOI}
\PrintBackRefs{\CurrentBib}

\bibitem [\protect \citeauthoryear {%
Gray%
}{%
Gray%
}{%
{\protect \APACyear {1984}}%
}]{%
vector_quantization}
\APACinsertmetastar {%
vector_quantization}%
\begin{APACrefauthors}%
Gray, R.%
\end{APACrefauthors}%
\unskip\
\newblock
\APACrefYearMonthDay{1984}{}{}.
\newblock
{\BBOQ}\APACrefatitle {Vector quantization} {Vector quantization}.{\BBCQ}
\newblock
\APACjournalVolNumPages{IEEE ASSP Magazine}{1}{2}{4-29}.
\newblock
\begin{APACrefDOI} \doi{10.1109/MASSP.1984.1162229} \end{APACrefDOI}
\PrintBackRefs{\CurrentBib}

\bibitem [\protect \citeauthoryear {%
Grimes%
\ \protect \BOthers {.}}{%
Grimes%
\ \protect \BOthers {.}}{%
{\protect \APACyear {2022}}%
}]{%
pyspedas}
\APACinsertmetastar {%
pyspedas}%
\begin{APACrefauthors}%
Grimes, E\BPBI W.%
, Harter, B.%
, Hatzigeorgiu, N.%
, Drozdov, A.%
, Lewis, J\BPBI W.%
, Angelopoulos, V.%
\BDBL {}Le~Contel, O.%
\end{APACrefauthors}%
\unskip\
\newblock
\APACrefYearMonthDay{2022}{}{}.
\newblock
{\BBOQ}\APACrefatitle {The Space Physics Environment Data Analysis System in Python} {The space physics environment data analysis system in python}.{\BBCQ}
\newblock
\APACjournalVolNumPages{Frontiers in Astronomy and Space Sciences}{9}{}{}.
\newblock
\begin{APACrefURL} \url{https://www.frontiersin.org/articles/10.3389/fspas.2022.1020815} \end{APACrefURL}
\newblock
\begin{APACrefDOI} \doi{10.3389/fspas.2022.1020815} \end{APACrefDOI}
\PrintBackRefs{\CurrentBib}

\bibitem [\protect \citeauthoryear {%
Gringauz%
, Bezrukikh%
, Ozerov%
\BCBL {}\ \BBA {} Rybchinskii%
}{%
Gringauz%
\ \protect \BOthers {.}}{%
{\protect \APACyear {1962}}%
}]{%
first_SW_measurement_Soviet}
\APACinsertmetastar {%
first_SW_measurement_Soviet}%
\begin{APACrefauthors}%
Gringauz, K.%
, Bezrukikh, V.%
, Ozerov, V.%
\BCBL {}\ \BBA {} Rybchinskii, R.%
\end{APACrefauthors}%
\unskip\
\newblock
\APACrefYearMonthDay{1962}{}{}.
\newblock
{\BBOQ}\APACrefatitle {The study of interplanetary ionized gas, high-energy electrons and corpuscular radiation of the sun, employing three-electrode charged particle traps on the second Soviet space rocket} {The study of interplanetary ionized gas, high-energy electrons and corpuscular radiation of the sun, employing three-electrode charged particle traps on the second soviet space rocket}.{\BBCQ}
\newblock
\APACjournalVolNumPages{Planetary and Space Science}{9}{3}{103-107}.
\newblock
\begin{APACrefURL} \url{https://www.sciencedirect.com/science/article/pii/0032063362901800} \end{APACrefURL}
\newblock
\begin{APACrefDOI} \doi{https://doi.org/10.1016/0032-0633(62)90180-0} \end{APACrefDOI}
\PrintBackRefs{\CurrentBib}

\bibitem [\protect \citeauthoryear {%
Harris%
\ \protect \BOthers {.}}{%
Harris%
\ \protect \BOthers {.}}{%
{\protect \APACyear {2020}}%
}]{%
numpy}
\APACinsertmetastar {%
numpy}%
\begin{APACrefauthors}%
Harris, C\BPBI R.%
, Millman, K\BPBI J.%
, van~der Walt, S\BPBI J.%
, Gommers, R.%
, Virtanen, P.%
, Cournapeau, D.%
\BDBL {}Oliphant, T\BPBI E.%
\end{APACrefauthors}%
\unskip\
\newblock
\APACrefYearMonthDay{2020}{{\APACmonth{09}}}{}.
\newblock
{\BBOQ}\APACrefatitle {Array programming with {NumPy}} {Array programming with {NumPy}}.{\BBCQ}
\newblock
\APACjournalVolNumPages{Nature}{585}{7825}{357--362}.
\newblock
\begin{APACrefURL} \url{https://doi.org/10.1038/s41586-020-2649-2} \end{APACrefURL}
\newblock
\begin{APACrefDOI} \doi{10.1038/s41586-020-2649-2} \end{APACrefDOI}
\PrintBackRefs{\CurrentBib}

\bibitem [\protect \citeauthoryear {%
Horn%
, Pack%
\BCBL {}\ \BBA {} Rieger%
}{%
Horn%
\ \protect \BOthers {.}}{%
{\protect \APACyear {2020}}%
}]{%
autofeat}
\APACinsertmetastar {%
autofeat}%
\begin{APACrefauthors}%
Horn, F.%
, Pack, R.%
\BCBL {}\ \BBA {} Rieger, M.%
\end{APACrefauthors}%
\unskip\
\newblock
\APACrefYearMonthDay{2020}{}{}.
\newblock
{\BBOQ}\APACrefatitle {The autofeat Python Library for Automated Feature Engineering and Selection} {The autofeat python library for automated feature engineering and selection}.{\BBCQ}
\newblock
\BIn{} P.~Cellier\ \BBA {} K.~Driessens\ (\BEDS), \APACrefbtitle {Machine Learning and Knowledge Discovery in Databases} {Machine learning and knowledge discovery in databases}\ (\BPGS\ 111--120).
\newblock
\APACaddressPublisher{Cham}{Springer International Publishing}.
\PrintBackRefs{\CurrentBib}

\bibitem [\protect \citeauthoryear {%
Innocenti%
\ \protect \BOthers {.}}{%
Innocenti%
\ \protect \BOthers {.}}{%
{\protect \APACyear {2021}}%
}]{%
som_openggcm}
\APACinsertmetastar {%
som_openggcm}%
\begin{APACrefauthors}%
Innocenti, M\BPBI E.%
, Amaya, J.%
, Raeder, J.%
, Dupuis, R.%
, Ferdousi, B.%
\BCBL {}\ \BBA {} Lapenta, G.%
\end{APACrefauthors}%
\unskip\
\newblock
\APACrefYearMonthDay{2021}{}{}.
\newblock
{\BBOQ}\APACrefatitle {Unsupervised classification of simulated magnetospheric regions} {Unsupervised classification of simulated magnetospheric regions}.{\BBCQ}
\newblock
\APACjournalVolNumPages{Annales Geophysicae}{39}{5}{861--881}.
\newblock
\begin{APACrefURL} \url{https://angeo.copernicus.org/articles/39/861/2021/} \end{APACrefURL}
\newblock
\begin{APACrefDOI} \doi{10.5194/angeo-39-861-2021} \end{APACrefDOI}
\PrintBackRefs{\CurrentBib}

\bibitem [\protect \citeauthoryear {%
Jolliffe%
}{%
Jolliffe%
}{%
{\protect \APACyear {2011}}%
}]{%
pca}
\APACinsertmetastar {%
pca}%
\begin{APACrefauthors}%
Jolliffe, I.%
\end{APACrefauthors}%
\unskip\
\newblock
\APACrefYearMonthDay{2011}{}{}.
\newblock
{\BBOQ}\APACrefatitle {Principal Component Analysis} {Principal component analysis}.{\BBCQ}
\newblock
\BIn{} M.~Lovric\ (\BED), \APACrefbtitle {International Encyclopedia of Statistical Science} {International encyclopedia of statistical science}\ (\BPGS\ 1094--1096).
\newblock
\APACaddressPublisher{Berlin, Heidelberg}{Springer Berlin Heidelberg}.
\newblock
\begin{APACrefURL} \url{https://doi.org/10.1007/978-3-642-04898-2_455} \end{APACrefURL}
\newblock
\begin{APACrefDOI} \doi{10.1007/978-3-642-04898-2_455} \end{APACrefDOI}
\PrintBackRefs{\CurrentBib}

\bibitem [\protect \citeauthoryear {%
Kohler%
\ \BBA {} Luniak%
}{%
Kohler%
\ \BBA {} Luniak%
}{%
{\protect \APACyear {2005}}%
}]{%
pca_biplot}
\APACinsertmetastar {%
pca_biplot}%
\begin{APACrefauthors}%
Kohler, U.%
\BCBT {}\ \BBA {} Luniak, M.%
\end{APACrefauthors}%
\unskip\
\newblock
\APACrefYearMonthDay{2005}{}{}.
\newblock
{\BBOQ}\APACrefatitle {Data Inspection using Biplots} {Data inspection using biplots}.{\BBCQ}
\newblock
\APACjournalVolNumPages{The Stata Journal}{5}{2}{208-223}.
\newblock
\begin{APACrefURL} \url{https://doi.org/10.1177/1536867X0500500206} \end{APACrefURL}
\newblock
\begin{APACrefDOI} \doi{10.1177/1536867X0500500206} \end{APACrefDOI}
\PrintBackRefs{\CurrentBib}

\bibitem [\protect \citeauthoryear {%
Kohonen%
}{%
Kohonen%
}{%
{\protect \APACyear {1982}}%
}]{%
Kohonen_original_som}
\APACinsertmetastar {%
Kohonen_original_som}%
\begin{APACrefauthors}%
Kohonen, T.%
\end{APACrefauthors}%
\unskip\
\newblock
\APACrefYearMonthDay{1982}{Jan}{01}.
\newblock
{\BBOQ}\APACrefatitle {Self-organized formation of topologically correct feature maps} {Self-organized formation of topologically correct feature maps}.{\BBCQ}
\newblock
\APACjournalVolNumPages{Biological Cybernetics}{43}{1}{59-69}.
\newblock
\begin{APACrefURL} \url{https://doi.org/10.1007/BF00337288} \end{APACrefURL}
\newblock
\begin{APACrefDOI} \doi{10.1007/BF00337288} \end{APACrefDOI}
\PrintBackRefs{\CurrentBib}

\bibitem [\protect \citeauthoryear {%
Kohonen%
}{%
Kohonen%
}{%
{\protect \APACyear {2014}}%
}]{%
Kohonen_matlab_som_intro}
\APACinsertmetastar {%
Kohonen_matlab_som_intro}%
\begin{APACrefauthors}%
Kohonen, T.%
\end{APACrefauthors}%
\unskip\
\newblock
\APACrefYear{2014}.
\newblock
\APACrefbtitle {MATLAB Implementations and Applications of the Self-Organizing Map} {Matlab implementations and applications of the self-organizing map}.
\newblock
\APACaddressPublisher{Unigrafia Oy, Helsinki, Finland}{}.
\PrintBackRefs{\CurrentBib}

\bibitem [\protect \citeauthoryear {%
Köhne%
, Boella%
\BCBL {}\ \BBA {} Innocenti%
}{%
Köhne%
\ \protect \BOthers {.}}{%
{\protect \APACyear {2023}}%
}]{%
fully_kinetic_sim_SOM}
\APACinsertmetastar {%
fully_kinetic_sim_SOM}%
\begin{APACrefauthors}%
Köhne, S.%
, Boella, E.%
\BCBL {}\ \BBA {} Innocenti, M\BPBI E.%
\end{APACrefauthors}%
\unskip\
\newblock
\APACrefYearMonthDay{2023}{}{}.
\newblock
{\BBOQ}\APACrefatitle {Unsupervised classification of fully kinetic simulations of plasmoid instability using self-organizing maps (SOMs)} {Unsupervised classification of fully kinetic simulations of plasmoid instability using self-organizing maps (soms)}.{\BBCQ}
\newblock
\APACjournalVolNumPages{Journal of Plasma Physics}{89}{3}{895890301}.
\newblock
\begin{APACrefDOI} \doi{10.1017/S0022377823000454} \end{APACrefDOI}
\PrintBackRefs{\CurrentBib}

\bibitem [\protect \citeauthoryear {%
Lloyd%
}{%
Lloyd%
}{%
{\protect \APACyear {1982}}%
}]{%
kmeans}
\APACinsertmetastar {%
kmeans}%
\begin{APACrefauthors}%
Lloyd, S.%
\end{APACrefauthors}%
\unskip\
\newblock
\APACrefYearMonthDay{1982}{}{}.
\newblock
{\BBOQ}\APACrefatitle {Least squares quantization in PCM} {Least squares quantization in pcm}.{\BBCQ}
\newblock
\APACjournalVolNumPages{IEEE Transactions on Information Theory}{28}{2}{129-137}.
\newblock
\begin{APACrefDOI} \doi{10.1109/TIT.1982.1056489} \end{APACrefDOI}
\PrintBackRefs{\CurrentBib}

\bibitem [\protect \citeauthoryear {%
Lucek%
\ \protect \BOthers {.}}{%
Lucek%
\ \protect \BOthers {.}}{%
{\protect \APACyear {2005}}%
}]{%
magsheath}
\APACinsertmetastar {%
magsheath}%
\begin{APACrefauthors}%
Lucek, E\BPBI A.%
, Constantinescu, D.%
, Goldstein, M\BPBI L.%
, Pickett, J.%
, Pin{\c{c}}on, J\BPBI L.%
, Sahraoui, F.%
\BDBL {}Walker, S\BPBI N.%
\end{APACrefauthors}%
\unskip\
\newblock
\APACrefYearMonthDay{2005}{Jun}{01}.
\newblock
{\BBOQ}\APACrefatitle {The Magnetosheath} {The magnetosheath}.{\BBCQ}
\newblock
\APACjournalVolNumPages{Space Science Reviews}{118}{1}{95-152}.
\newblock
\begin{APACrefURL} \url{https://doi.org/10.1007/s11214-005-3825-2} \end{APACrefURL}
\newblock
\begin{APACrefDOI} \doi{10.1007/s11214-005-3825-2} \end{APACrefDOI}
\PrintBackRefs{\CurrentBib}

\bibitem [\protect \citeauthoryear {%
Mancini%
, Ritacco%
, Lanciano%
\BCBL {}\ \BBA {} Cucinotta%
}{%
Mancini%
\ \protect \BOthers {.}}{%
{\protect \APACyear {2020}}%
}]{%
xpysom}
\APACinsertmetastar {%
xpysom}%
\begin{APACrefauthors}%
Mancini, R.%
, Ritacco, A.%
, Lanciano, G.%
\BCBL {}\ \BBA {} Cucinotta, T.%
\end{APACrefauthors}%
\unskip\
\newblock
\APACrefYearMonthDay{2020}{}{}.
\newblock
{\BBOQ}\APACrefatitle {XPySom: High-Performance Self-Organizing Maps} {Xpysom: High-performance self-organizing maps}.{\BBCQ}
\newblock
\BIn{} \APACrefbtitle {2020 IEEE 32nd International Symposium on Computer Architecture and High Performance Computing (SBAC-PAD)} {2020 ieee 32nd international symposium on computer architecture and high performance computing (sbac-pad)}\ (\BPG~209-216).
\newblock
\begin{APACrefDOI} \doi{10.1109/SBAC-PAD49847.2020.00037} \end{APACrefDOI}
\PrintBackRefs{\CurrentBib}

\bibitem [\protect \citeauthoryear {%
McFadden%
\ \protect \BOthers {.}}{%
McFadden%
\ \protect \BOthers {.}}{%
{\protect \APACyear {2008}}%
}]{%
McFadden}
\APACinsertmetastar {%
McFadden}%
\begin{APACrefauthors}%
McFadden, J\BPBI P.%
, Carlson, C\BPBI W.%
, Larson, D.%
, Ludlam, M.%
, Abiad, R.%
, Elliott, B.%
\BDBL {}Angelopoulos, V.%
\end{APACrefauthors}%
\unskip\
\newblock
\APACrefYearMonthDay{2008}{Dec}{01}.
\newblock
{\BBOQ}\APACrefatitle {The THEMIS ESA Plasma Instrument and In-flight Calibration} {The themis esa plasma instrument and in-flight calibration}.{\BBCQ}
\newblock
\APACjournalVolNumPages{Space Science Reviews}{141}{1}{277-302}.
\newblock
\begin{APACrefURL} \url{https://doi.org/10.1007/s11214-008-9440-2} \end{APACrefURL}
\newblock
\begin{APACrefDOI} \doi{10.1007/s11214-008-9440-2} \end{APACrefDOI}
\PrintBackRefs{\CurrentBib}

\bibitem [\protect \citeauthoryear {%
McKinney%
}{%
McKinney%
}{%
{\protect \APACyear {2010}}%
}]{%
pandas}
\APACinsertmetastar {%
pandas}%
\begin{APACrefauthors}%
McKinney, W.%
\end{APACrefauthors}%
\unskip\
\newblock
\APACrefYearMonthDay{2010}{}{}.
\newblock
{\BBOQ}\APACrefatitle {Data Structures for Statistical Computing in Python} {Data structures for statistical computing in python}.{\BBCQ}
\newblock
\BIn{} S.~{van der Walt}\ \BBA {} J.~Millman\ (\BEDS), \APACrefbtitle {Proceedings of the 9th Python in Science Conference} {Proceedings of the 9th python in science conference}\ (\BPGS\ 56--61).
\newblock
\APACaddressPublisher{}{SciPy}.
\newblock
\begin{APACrefDOI} \doi{10.25080/Majora-92bf1922-00a} \end{APACrefDOI}
\PrintBackRefs{\CurrentBib}

\bibitem [\protect \citeauthoryear {%
Müllner%
}{%
Müllner%
}{%
{\protect \APACyear {2013}}%
}]{%
hierarch_agg_clustering_methods}
\APACinsertmetastar {%
hierarch_agg_clustering_methods}%
\begin{APACrefauthors}%
Müllner, D.%
\end{APACrefauthors}%
\unskip\
\newblock
\APACrefYearMonthDay{2013}{}{}.
\newblock
{\BBOQ}\APACrefatitle {fastcluster: Fast Hierarchical, Agglomerative Clustering Routines for R and Python} {fastcluster: Fast hierarchical, agglomerative clustering routines for r and python}.{\BBCQ}
\newblock
\APACjournalVolNumPages{Journal of Statistical Software}{53}{9}{1–18}.
\newblock
\begin{APACrefURL} \url{https://www.jstatsoft.org/index.php/jss/article/view/v053i09} \end{APACrefURL}
\newblock
\begin{APACrefDOI} \doi{10.18637/jss.v053.i09} \end{APACrefDOI}
\PrintBackRefs{\CurrentBib}

\bibitem [\protect \citeauthoryear {%
Nguyen%
\ \protect \BOthers {.}}{%
Nguyen%
\ \protect \BOthers {.}}{%
{\protect \APACyear {2022}}%
}]{%
nguyen_ml_method}
\APACinsertmetastar {%
nguyen_ml_method}%
\begin{APACrefauthors}%
Nguyen, G.%
, Aunai, N.%
, Michotte~de Welle, B.%
, Jeandet, A.%
, Lavraud, B.%
\BCBL {}\ \BBA {} Fontaine, D.%
\end{APACrefauthors}%
\unskip\
\newblock
\APACrefYearMonthDay{2022}{}{}.
\newblock
{\BBOQ}\APACrefatitle {Massive Multi-Mission Statistical Study and Analytical Modeling of the Earth's Magnetopause: 1. A Gradient Boosting Based Automatic Detection of Near-Earth Regions} {Massive multi-mission statistical study and analytical modeling of the earth's magnetopause: 1. a gradient boosting based automatic detection of near-earth regions}.{\BBCQ}
\newblock
\APACjournalVolNumPages{Journal of Geophysical Research: Space Physics}{127}{1}{e2021JA029773}.
\newblock
\begin{APACrefURL} \url{https://agupubs.onlinelibrary.wiley.com/doi/abs/10.1029/2021JA029773} \end{APACrefURL}
\newblock
\APACrefnote{e2021JA029773 2021JA029773}
\newblock
\begin{APACrefDOI} \doi{https://doi.org/10.1029/2021JA029773} \end{APACrefDOI}
\PrintBackRefs{\CurrentBib}

\bibitem [\protect \citeauthoryear {%
Nielsen%
}{%
Nielsen%
}{%
{\protect \APACyear {2016}}%
}]{%
hierarchical_clustering_overview}
\APACinsertmetastar {%
hierarchical_clustering_overview}%
\begin{APACrefauthors}%
Nielsen, F.%
\end{APACrefauthors}%
\unskip\
\newblock
\APACrefYearMonthDay{2016}{}{}.
\newblock
{\BBOQ}\APACrefatitle {Hierarchical Clustering} {Hierarchical clustering}.{\BBCQ}
\newblock
\BIn{} \APACrefbtitle {Introduction to HPC with MPI for Data Science} {Introduction to hpc with mpi for data science}\ (\BPGS\ 195--211).
\newblock
\APACaddressPublisher{Cham}{Springer International Publishing}.
\newblock
\begin{APACrefURL} \url{https://doi.org/10.1007/978-3-319-21903-5_8} \end{APACrefURL}
\newblock
\begin{APACrefDOI} \doi{10.1007/978-3-319-21903-5_8} \end{APACrefDOI}
\PrintBackRefs{\CurrentBib}

\bibitem [\protect \citeauthoryear {%
Olshevsky%
\ \protect \BOthers {.}}{%
Olshevsky%
\ \protect \BOthers {.}}{%
{\protect \APACyear {2021}}%
}]{%
mms_cnn_olshevsky}
\APACinsertmetastar {%
mms_cnn_olshevsky}%
\begin{APACrefauthors}%
Olshevsky, V.%
, Khotyaintsev, Y\BPBI V.%
, Lalti, A.%
, Divin, A.%
, Delzanno, G\BPBI L.%
, Anderzén, S.%
\BDBL {}Markidis, S.%
\end{APACrefauthors}%
\unskip\
\newblock
\APACrefYearMonthDay{2021}{}{}.
\newblock
{\BBOQ}\APACrefatitle {Automated Classification of Plasma Regions Using 3D Particle Energy Distributions} {Automated classification of plasma regions using 3d particle energy distributions}.{\BBCQ}
\newblock
\APACjournalVolNumPages{Journal of Geophysical Research: Space Physics}{126}{10}{e2021JA029620}.
\newblock
\begin{APACrefURL} \url{https://agupubs.onlinelibrary.wiley.com/doi/abs/10.1029/2021JA029620} \end{APACrefURL}
\newblock
\APACrefnote{e2021JA029620 2021JA029620}
\newblock
\begin{APACrefDOI} \doi{https://doi.org/10.1029/2021JA029620} \end{APACrefDOI}
\PrintBackRefs{\CurrentBib}

\bibitem [\protect \citeauthoryear {%
Pedregosa%
\ \protect \BOthers {.}}{%
Pedregosa%
\ \protect \BOthers {.}}{%
{\protect \APACyear {2011}}%
}]{%
scikit}
\APACinsertmetastar {%
scikit}%
\begin{APACrefauthors}%
Pedregosa, F.%
, Varoquaux, G.%
, Gramfort, A.%
, Michel, V.%
, Thirion, B.%
, Grisel, O.%
\BDBL {}others%
\end{APACrefauthors}%
\unskip\
\newblock
\APACrefYearMonthDay{2011}{}{}.
\newblock
{\BBOQ}\APACrefatitle {Scikit-learn: Machine learning in Python} {Scikit-learn: Machine learning in python}.{\BBCQ}
\newblock
\APACjournalVolNumPages{Journal of machine learning research}{12}{Oct}{2825--2830}.
\PrintBackRefs{\CurrentBib}

\bibitem [\protect \citeauthoryear {%
Pollock%
\ \protect \BOthers {.}}{%
Pollock%
\ \protect \BOthers {.}}{%
{\protect \APACyear {2016}}%
}]{%
MMS_FPI}
\APACinsertmetastar {%
MMS_FPI}%
\begin{APACrefauthors}%
Pollock, C.%
, Moore, T.%
, Jacques, A.%
, Burch, J.%
, Gliese, U.%
, Saito, Y.%
\BDBL {}Zeuch, M.%
\end{APACrefauthors}%
\unskip\
\newblock
\APACrefYearMonthDay{2016}{Mar}{01}.
\newblock
{\BBOQ}\APACrefatitle {Fast Plasma Investigation for Magnetospheric Multiscale} {Fast plasma investigation for magnetospheric multiscale}.{\BBCQ}
\newblock
\APACjournalVolNumPages{Space Science Reviews}{199}{1}{331-406}.
\newblock
\begin{APACrefURL} \url{https://doi.org/10.1007/s11214-016-0245-4} \end{APACrefURL}
\newblock
\begin{APACrefDOI} \doi{10.1007/s11214-016-0245-4} \end{APACrefDOI}
\PrintBackRefs{\CurrentBib}

\bibitem [\protect \citeauthoryear {%
Russell%
\ \protect \BOthers {.}}{%
Russell%
\ \protect \BOthers {.}}{%
{\protect \APACyear {2016}}%
}]{%
MMS_FGM}
\APACinsertmetastar {%
MMS_FGM}%
\begin{APACrefauthors}%
Russell, C\BPBI T.%
, Anderson, B\BPBI J.%
, Baumjohann, W.%
, Bromund, K\BPBI R.%
, Dearborn, D.%
, Fischer, D.%
\BDBL {}Richter, I.%
\end{APACrefauthors}%
\unskip\
\newblock
\APACrefYearMonthDay{2016}{Mar}{01}.
\newblock
{\BBOQ}\APACrefatitle {The Magnetospheric Multiscale Magnetometers} {The magnetospheric multiscale magnetometers}.{\BBCQ}
\newblock
\APACjournalVolNumPages{Space Science Reviews}{199}{1}{189-256}.
\newblock
\begin{APACrefURL} \url{https://doi.org/10.1007/s11214-014-0057-3} \end{APACrefURL}
\newblock
\begin{APACrefDOI} \doi{10.1007/s11214-014-0057-3} \end{APACrefDOI}
\PrintBackRefs{\CurrentBib}

\bibitem [\protect \citeauthoryear {%
Sakia%
}{%
Sakia%
}{%
{\protect \APACyear {2018}}%
}]{%
box_cox}
\APACinsertmetastar {%
box_cox}%
\begin{APACrefauthors}%
Sakia, R\BPBI M.%
\end{APACrefauthors}%
\unskip\
\newblock
\APACrefYearMonthDay{2018}{12}{}.
\newblock
{\BBOQ}\APACrefatitle {{The Box-Cox Transformation Technique: A Review}} {{The Box-Cox Transformation Technique: A Review}}.{\BBCQ}
\newblock
\APACjournalVolNumPages{Journal of the Royal Statistical Society Series D: The Statistician}{41}{2}{169-178}.
\newblock
\begin{APACrefURL} \url{https://doi.org/10.2307/2348250} \end{APACrefURL}
\newblock
\begin{APACrefDOI} \doi{10.2307/2348250} \end{APACrefDOI}
\PrintBackRefs{\CurrentBib}

\bibitem [\protect \citeauthoryear {%
Shue%
\ \protect \BOthers {.}}{%
Shue%
\ \protect \BOthers {.}}{%
{\protect \APACyear {1998}}%
}]{%
shue_mp}
\APACinsertmetastar {%
shue_mp}%
\begin{APACrefauthors}%
Shue, J\BHBI H.%
, Song, P.%
, Russell, C\BPBI T.%
, Steinberg, J\BPBI T.%
, Chao, J\BPBI K.%
, Zastenker, G.%
\BDBL {}Kawano, H.%
\end{APACrefauthors}%
\unskip\
\newblock
\APACrefYearMonthDay{1998}{}{}.
\newblock
{\BBOQ}\APACrefatitle {Magnetopause location under extreme solar wind conditions} {Magnetopause location under extreme solar wind conditions}.{\BBCQ}
\newblock
\APACjournalVolNumPages{Journal of Geophysical Research: Space Physics}{103}{A8}{17691-17700}.
\newblock
\begin{APACrefURL} \url{https://agupubs.onlinelibrary.wiley.com/doi/abs/10.1029/98JA01103} \end{APACrefURL}
\newblock
\begin{APACrefDOI} \doi{https://doi.org/10.1029/98JA01103} \end{APACrefDOI}
\PrintBackRefs{\CurrentBib}

\bibitem [\protect \citeauthoryear {%
Smith%
\ \protect \BOthers {.}}{%
Smith%
\ \protect \BOthers {.}}{%
{\protect \APACyear {2020}}%
}]{%
ssc_ml}
\APACinsertmetastar {%
ssc_ml}%
\begin{APACrefauthors}%
Smith, A\BPBI W.%
, Rae, I\BPBI J.%
, Forsyth, C.%
, Oliveira, D\BPBI M.%
, Freeman, M\BPBI P.%
\BCBL {}\ \BBA {} Jackson, D\BPBI R.%
\end{APACrefauthors}%
\unskip\
\newblock
\APACrefYearMonthDay{2020}{}{}.
\newblock
{\BBOQ}\APACrefatitle {Probabilistic Forecasts of Storm Sudden Commencements From Interplanetary Shocks Using Machine Learning} {Probabilistic forecasts of storm sudden commencements from interplanetary shocks using machine learning}.{\BBCQ}
\newblock
\APACjournalVolNumPages{Space Weather}{18}{11}{e2020SW002603}.
\newblock
\begin{APACrefURL} \url{https://agupubs.onlinelibrary.wiley.com/doi/abs/10.1029/2020SW002603} \end{APACrefURL}
\newblock
\APACrefnote{e2020SW002603 10.1029/2020SW002603}
\newblock
\begin{APACrefDOI} \doi{https://doi.org/10.1029/2020SW002603} \end{APACrefDOI}
\PrintBackRefs{\CurrentBib}

\bibitem [\protect \citeauthoryear {%
Snyder%
, Neugebauer%
\BCBL {}\ \BBA {} Rao%
}{%
Snyder%
\ \protect \BOthers {.}}{%
{\protect \APACyear {1963}}%
}]{%
first_SW_measurement_American}
\APACinsertmetastar {%
first_SW_measurement_American}%
\begin{APACrefauthors}%
Snyder, C\BPBI W.%
, Neugebauer, M.%
\BCBL {}\ \BBA {} Rao, U\BPBI R.%
\end{APACrefauthors}%
\unskip\
\newblock
\APACrefYearMonthDay{1963}{}{}.
\newblock
{\BBOQ}\APACrefatitle {The solar wind velocity and its correlation with cosmic-ray variations and with solar and geomagnetic activity} {The solar wind velocity and its correlation with cosmic-ray variations and with solar and geomagnetic activity}.{\BBCQ}
\newblock
\APACjournalVolNumPages{Journal of Geophysical Research (1896-1977)}{68}{24}{6361-6370}.
\newblock
\begin{APACrefURL} \url{https://agupubs.onlinelibrary.wiley.com/doi/abs/10.1029/JZ068i024p06361} \end{APACrefURL}
\newblock
\begin{APACrefDOI} \doi{https://doi.org/10.1029/JZ068i024p06361} \end{APACrefDOI}
\PrintBackRefs{\CurrentBib}

\bibitem [\protect \citeauthoryear {%
Vettigli%
}{%
Vettigli%
}{%
{\protect \APACyear {2018}}%
}]{%
minisom}
\APACinsertmetastar {%
minisom}%
\begin{APACrefauthors}%
Vettigli, G.%
\end{APACrefauthors}%
\unskip\
\newblock
\APACrefYearMonthDay{2018}{}{}.
\newblock
\APACrefbtitle {MiniSom: minimalistic and NumPy-based implementation of the Self Organizing Map.} {Minisom: minimalistic and numpy-based implementation of the self organizing map.}
\newblock
\begin{APACrefURL} \url{https://github.com/JustGlowing/minisom/} \end{APACrefURL}
\PrintBackRefs{\CurrentBib}

\bibitem [\protect \citeauthoryear {%
Virtanen%
\ \protect \BOthers {.}}{%
Virtanen%
\ \protect \BOthers {.}}{%
{\protect \APACyear {2020}}%
}]{%
scipy}
\APACinsertmetastar {%
scipy}%
\begin{APACrefauthors}%
Virtanen, P.%
, Gommers, R.%
, Oliphant, T\BPBI E.%
, Haberland, M.%
, Reddy, T.%
, Cournapeau, D.%
\BDBL {}{SciPy 1.0 Contributors}%
\end{APACrefauthors}%
\unskip\
\newblock
\APACrefYearMonthDay{2020}{}{}.
\newblock
{\BBOQ}\APACrefatitle {{{SciPy} 1.0: Fundamental Algorithms for Scientific Computing in Python}} {{{SciPy} 1.0: Fundamental Algorithms for Scientific Computing in Python}}.{\BBCQ}
\newblock
\APACjournalVolNumPages{Nature Methods}{17}{}{261--272}.
\newblock
\begin{APACrefDOI} \doi{10.1038/s41592-019-0686-2} \end{APACrefDOI}
\PrintBackRefs{\CurrentBib}

\bibitem [\protect \citeauthoryear {%
Willis%
}{%
Willis%
}{%
{\protect \APACyear {1971}}%
}]{%
magnetopause}
\APACinsertmetastar {%
magnetopause}%
\begin{APACrefauthors}%
Willis, D\BPBI M.%
\end{APACrefauthors}%
\unskip\
\newblock
\APACrefYearMonthDay{1971}{}{}.
\newblock
{\BBOQ}\APACrefatitle {Structure of the magnetopause} {Structure of the magnetopause}.{\BBCQ}
\newblock
\APACjournalVolNumPages{Reviews of Geophysics}{9}{4}{953-985}.
\newblock
\begin{APACrefURL} \url{https://agupubs.onlinelibrary.wiley.com/doi/abs/10.1029/RG009i004p00953} \end{APACrefURL}
\newblock
\begin{APACrefDOI} \doi{https://doi.org/10.1029/RG009i004p00953} \end{APACrefDOI}
\PrintBackRefs{\CurrentBib}

\bibitem [\protect \citeauthoryear {%
Wittek%
, Gao%
, Lim%
\BCBL {}\ \BBA {} Zhao%
}{%
Wittek%
\ \protect \BOthers {.}}{%
{\protect \APACyear {2017}}%
}]{%
somoclu}
\APACinsertmetastar {%
somoclu}%
\begin{APACrefauthors}%
Wittek, P.%
, Gao, S\BPBI C.%
, Lim, I\BPBI S.%
\BCBL {}\ \BBA {} Zhao, L.%
\end{APACrefauthors}%
\unskip\
\newblock
\APACrefYearMonthDay{2017}{}{}.
\newblock
{\BBOQ}\APACrefatitle {somoclu: An Efficient Parallel Library for Self-Organizing Maps} {somoclu: An efficient parallel library for self-organizing maps}.{\BBCQ}
\newblock
\APACjournalVolNumPages{Journal of Statistical Software}{78}{9}{1–21}.
\newblock
\begin{APACrefURL} \url{https://www.jstatsoft.org/index.php/jss/article/view/v078i09} \end{APACrefURL}
\newblock
\begin{APACrefDOI} \doi{10.18637/jss.v078.i09} \end{APACrefDOI}
\PrintBackRefs{\CurrentBib}

\end{thebibliography}

%
%
%
%
%

\end{document}